\documentclass[aps,twocolumn,groupedaddress,showpacs,prb,floatfix]{revtex4}
\usepackage{graphicx,rotating,subfigure,amsmath,amsfonts,amssymb,delarray}
\renewcommand{\vec}[1]{\boldsymbol #1}
\newcommand{\e}{\text{e}}
\newcommand{\im}{\text{i}}
\def\l{\left}
\def\r{\right}
\def\12{\frac{1}{2}}
\def\nn{\nonumber}
\newcommand{\rep}[2]{{#1}^{({#2})}}

\begin{document}
\bibliographystyle{apsrev}


\title{Doping a Mott insulator with orbital degrees of freedom}


\author{J. Sirker}
\email[]{j.sirker@fkf.mpg.de}
\affiliation{Max-Planck-Institut f\"ur Festk\"orperforschung, Heisenbergstr.~1,
  70569 Stuttgart, Germany}
\author{J. Damerau}
\email[]{damerau@physik.uni-wuppertal.de}
\affiliation{Bergische Universit\"at Wuppertal, Fachbereich Physik, 42097
  Wuppertal, Germany}
\author{A. Kl\"umper}
\email[]{kluemper@physik.uni-wuppertal.de}
\affiliation{Bergische Universit\"at Wuppertal, Fachbereich Physik, 42097
  Wuppertal, Germany}

\date{\today}

\begin{abstract}
  We study the effects of hole doping on one-dimensional Mott insulators with
  orbital degrees of freedom. We describe the system in terms of a generalized
  $t-J$ model. At a specific point in parameter space the model becomes
  integrable in analogy to the one-band supersymmetric $t-J$ model. We use
  the Bethe ansatz to derive a set of nonlinear integral equations which
  allow us to study the thermodynamics exactly. Moving away from this special
  point in parameter space we use the density-matrix renormalization group
  applied to transfer matrices to study the evolution of various phases of the
  undoped system with doping and temperature. Finally, we study a
  one-dimensional version of a realistic model for cubic titanates which
  includes the anisotropy of the orbital sector due to Hund's coupling. We
  find a transition from a phase with antiferromagnetically correlated spins
  to a phase where the spins are fully ferromagnetically polarized, a strong
  tendency towards phase separation at large Hund's coupling, as well as the
  possibility of an instability towards triplet superconductivity.
\end{abstract}
\pacs{71.10.Fd, 05.70.-a, 05.10.Cc}

\maketitle

\section{Introduction}
\label{Intro}
In many transition metal oxides, different orbital configurations are close in
energy or even degenerate. Small changes in temperature or pressure can
therefore lead to a complete rearrangement of the electron clouds which in
turn also strongly influences the magnetic and the transport properties. Such
orbital degrees of freedom play an important role, for example, in the
manganites, titanates, vanadates, and
ruthenates.\cite{TokuraNagaosa,Khaliullin_Rev} Quite common for these
transition metal oxides is the perovskite crystal structure where each
transition metal ion is surrounded by an octahedron of oxygen ions. For a $3d$
transition metal ion, the cubic crystal field then splits the fivefold orbital
degeneracy into threefold degenerate $t_{2g}$ orbitals and twofold degenerate
$e_g$ orbitals. Here we want to concentrate first on the case where we have
one electron per site in the $t_{2g}$ orbitals with the $e_g$ orbitals being
inactive (fully occupied or empty). Because the onsite Coulomb interactions
are large the system is a Mott insulator in this case. The strongly
anisotropic shape of the $t_{2g}$ orbitals means that the direction an
electron can move to create a virtually excited state depends on the orbital
it is sitting in.  More precisely, hopping is only possible between orbitals
of the same kind, and along a particular crystal axis only two out of the
three $t_{2g}$ orbitals are active. This can lead to a (dynamical) lowering of
the effective dimensionality of the system in various ways: Conventional
orbital ordering can restrict the hopping to one-dimensional chains which can
then show typical one-dimensional phenomena like a Haldane gap \cite{LeePark}
or a Peierls effect.\cite{KhomskiiMizokawa} More unconventional mechanisms
like an orbital-driven Peierls effect\cite{KeimerSirker,SirkerKhaliullin} or
spin-orbital nematic states \cite{Khaliullin_Rev} might also render the system
quasi one dimensional. In the latter cases, however, there will still be a
twofold orbital degeneracy. A simple Hamiltonian capturing the essential
physics is then given by
\begin{equation}
\label{intro1}
H = 2J\sum_j \l(\vec{S}_j\vec{S}_{j+1} + x\r)\l(\vec{\tau}_j\vec{\tau}_{j+1} + y\r)
\end{equation}    
where $\vec{S}$ is an $S=1/2$ spin operator and $\vec{\tau}$ a $\tau=1/2$
orbital pseudospin describing the occupation of the two degenerate orbitals
active along the chain direction. $J = 4t^2/U$ is the magnetic superexchange
constant, $t$ the hopping amplitude, $U$ the onsite Coulomb repulsion, and
$x,y$ real numbers often treated as free parameters. From a microscopic
derivation of the effective model (\ref{intro1}) it follows, however, that
$x,y$ are determined by the Hund's rule coupling $J_H$ with $x=y=1/4$
corresponding to $J_H=0$. In addition, such a derivation shows that a finite
Hund's coupling does not only modify $x,y$ but also leads to an $xxz$-type
anisotropy of the orbital sector.\cite{Khaliullin_Rev} This anisotropy is
neglected in (\ref{intro1}).  We will come back to the relation between this
simple model and more realistic models in section \ref{titanates}.

The spin-orbital model (\ref{intro1}) has been intensely studied
\cite{LiMa,YamashitaShibata,PatiSingh,FrischmuthMila,ItoiQin,AzariaBoulat,PatiSinghKhomskii,SirkerSu4}
and a number of different phases depending on $x,y$ have been identified (see
e.g.~Refs.~\onlinecite{ItoiQin}, \onlinecite{ChenWang}). In general, the
model has a $SU(2)\times SU(2)$ symmetry and exhibits an additional $Z_2$
symmetry, interchanging spin and orbital degrees of freedom, if $x=y$. At the
special point $x=y=1/4$ the symmetry is enlarged even further to $SU(4)$. This
has to do with the fact that at this point the Hamiltonian is just a
permutation operator of states on neighboring sites. The model therefore
becomes a version of the Uimin-Sutherland model and is integrable by Bethe
ansatz.\cite{Sutherland}

In this work we want to study the effects of hole doping on the
spin-orbital model (\ref{intro1}). Because states with more than one
electron per site are effectively forbidden due to the strong Coulomb
repulsion $U$, hole doping of the Mott insulator (\ref{intro1})
naturally leads us to a generalized $t-J$ model
\begin{widetext}
\begin{equation}
\label{intro2}
H  = t\sum_j\sum_{\sigma,\tau}\mathcal{P}\l\{
c^\dagger_{j,\sigma,\tau}c_{j+1,\sigma,\tau} + h.c. \r\}\mathcal{P} 
+ 2J\sum_j \l\{\l(\vec{S}_j\vec{S}_{j+1} +
x\r)\l(\vec{\tau}_j\vec{\tau}_{j+1} + yn_jn_{j+1}\r)-\frac{n_jn_{j+1}}{4}\r\} \; . 
\end{equation}
\end{widetext}
Here $\mathcal{P}$ projects out the doubly occupied states, $\sigma$ is the
spin, and $\tau$ the orbital index. As for the one-band $t-J$ model it turns
out that there is a special point in parameter space $J/t =2$, $x=y=1/4$ where
the model is integrable by Bethe ansatz. Again, the symmetry is enlarged at
this point to $SU(4|1)$ (graded $SU(5)$ symmetry), the Hamiltonian is a
permutation operator of states on neighboring sites and falls into the
Uimin-Sutherland class of models.

In Sec.~\ref{integrable} we will investigate the thermodynamics of this model
at the integrable point with the help of the Bethe ansatz and the quantum
transfer matrix approach. Details of the Bethe ansatz calculation are
presented in appendix \ref{appA}. In Sec.~\ref{TMRG} we briefly introduce the
density-matrix renormalization group applied to transfer matrices (TMRG) which
we will use to study the thermodynamics of model (\ref{intro2}) numerically
away from the integrable point. We will test the accuracy of this method by
comparing with exact results at the integrable point. In section
\ref{xy-model} we will use the TMRG algorithm to study the evolution of
various phases of the undoped model (\ref{intro1}) with doping and
temperature. In Sec.~\ref{titanates} we finally consider a one-dimensional
version of a realistic model for cubic titanates which includes the $xxz$-type
anisotropy of the orbital sector due to Hund's coupling. We investigate the
phase transitions, possible tendencies towards phase separation as well as
superconducting instabilities as a function of the strength of Hund's
coupling. In Sec.~\ref{Conc} we present a short summary and our
conclusions.
\section{The integrable model}
\label{integrable}
The integrable $SU(4|1)$ model (\ref{intro2}) with $J/t =2$, $x=y=1/4$ has
already been studied by Schlottmann.
\cite{Schlottmann_SU5_1,Schlottmann_SU5_2} He derived the Bethe ansatz
equations and studied the ground state properties as well as the elementary
excitations. Kawakami later then derived the critical exponents of various
correlation functions. \cite{Kawakami_SUN} Here we want to concentrate on the
thermodynamics of this model.  Based on the quantum transfer matrix approach
we derive a set of nonlinear integral equations (NLIE) which then are
evaluated numerically to obtain various thermodynamic quantities. Details
about the derivation of the NLIE are given in appendix \ref{appA}.

Our results for the thermodynamics in the low temperature limit can be
connected to Schlottmann's and Kawakami's results for the elementary
excitations using conformal field theory. The $SU(4)$ spin-orbital model,
i.e., model (\ref{intro2}) with $n=1$, is known to belong to the universality
class of the $SU(4)_1$ Wess-Zumino-Witten (WZW)
models,\cite{Affleck_SU(N),ItoiQin} so that the central charge $c_{so} =3$.
Similar to the one-band supersymmetric $t-J$ model we expect that the critical
theory for the hole-doped model is a semidirect product of the spin-orbital
and the charge part so that the free energy at low temperatures is given by
\begin{equation}
\label{BA1}
f=e_0 -\frac{\pi}{6}\l(\frac{c_{so}}{v_{so}}+\frac{c_{c}}{v_{c}}\r)T^2 \; .
\end{equation}
Here $e_0$ is the ground state energy which can be calculated by Bethe
ansatz\cite{Schlottmann_SU5_2}, $v_{so}$ ($v_c$) are the velocities of the
elementary spin-orbital (charge) excitations, respectively, and $c_c=1$ the
central charge of the charge sector.



\begin{figure}
\includegraphics*[width=0.9\columnwidth]{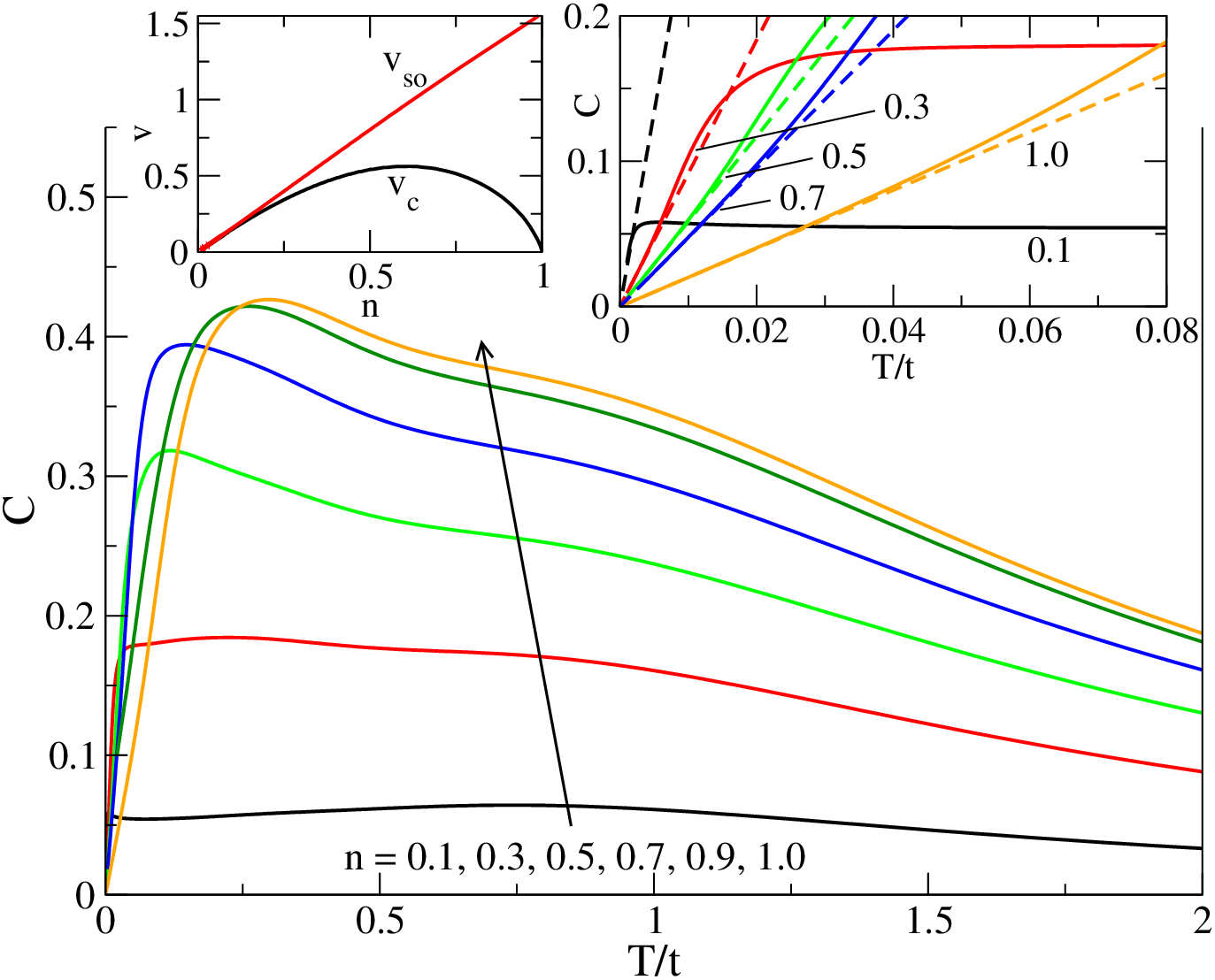}
\caption{(color online) Specific heat for the integrable model (\ref{intro2})
  with $J/t=2$ and $x=y=1/4$ calculated by BA. The left inset shows the
  spin-orbital ($v_{so}$) and the charge velocity ($v_c$) as a function of
  particle density $n$. The right inset compares the BA results (solid lines)
  with the low temperature asymptotics obtained from CFT (dashed lines), see
  Eq.~(\ref{BA2}).}
\label{BA.fig1}
\end{figure}

In Fig.~\ref{BA.fig1} we show the specific heat as a function of temperature
for various fillings. 
According to Eq.~(\ref{BA1}) the specific heat at low temperatures is
linear and determined by the elementary charge and spin-orbital excitations
\begin{equation}
\label{BA2}
C=-T\frac{\partial^2f}{\partial T^2}=\frac{\pi}{3}\l(\frac{c_{so}}{v_{so}}+\frac{c_{c}}{v_{c}}\r)T \; .
\end{equation}
As shown in the left inset of Fig.~\ref{BA.fig1} the velocities of the charge
and spin-orbital excitations go to zero for $n\to 0$ so that the slope of $C$
diverges in this limit. For $n\to 1$, on the other hand, only $v_c\to 0$
whereas $v_{so}\to J\pi/4$. This also leads to a diverging slope, however, the
charge excitations are quickly exhausted so that this behavior is only
visible at very low temperatures. At higher temperatures (but still $T\ll t$)
the slope then crosses over to the value given by the spin-orbital excitations
only. Finally, for $n=1$, we have $C=2T$ in the whole conformal regime. In the
low temperature limit, the BA results which we obtained by the quantum
transfer matrix approach indeed agree perfectly with the CFT result
(\ref{BA2}) using the velocities determined according to
Ref.~\onlinecite{Schlottmann_SU5_2} (see right inset of Fig.~\ref{BA.fig1}).

\begin{figure}
\includegraphics*[width=0.9\columnwidth]{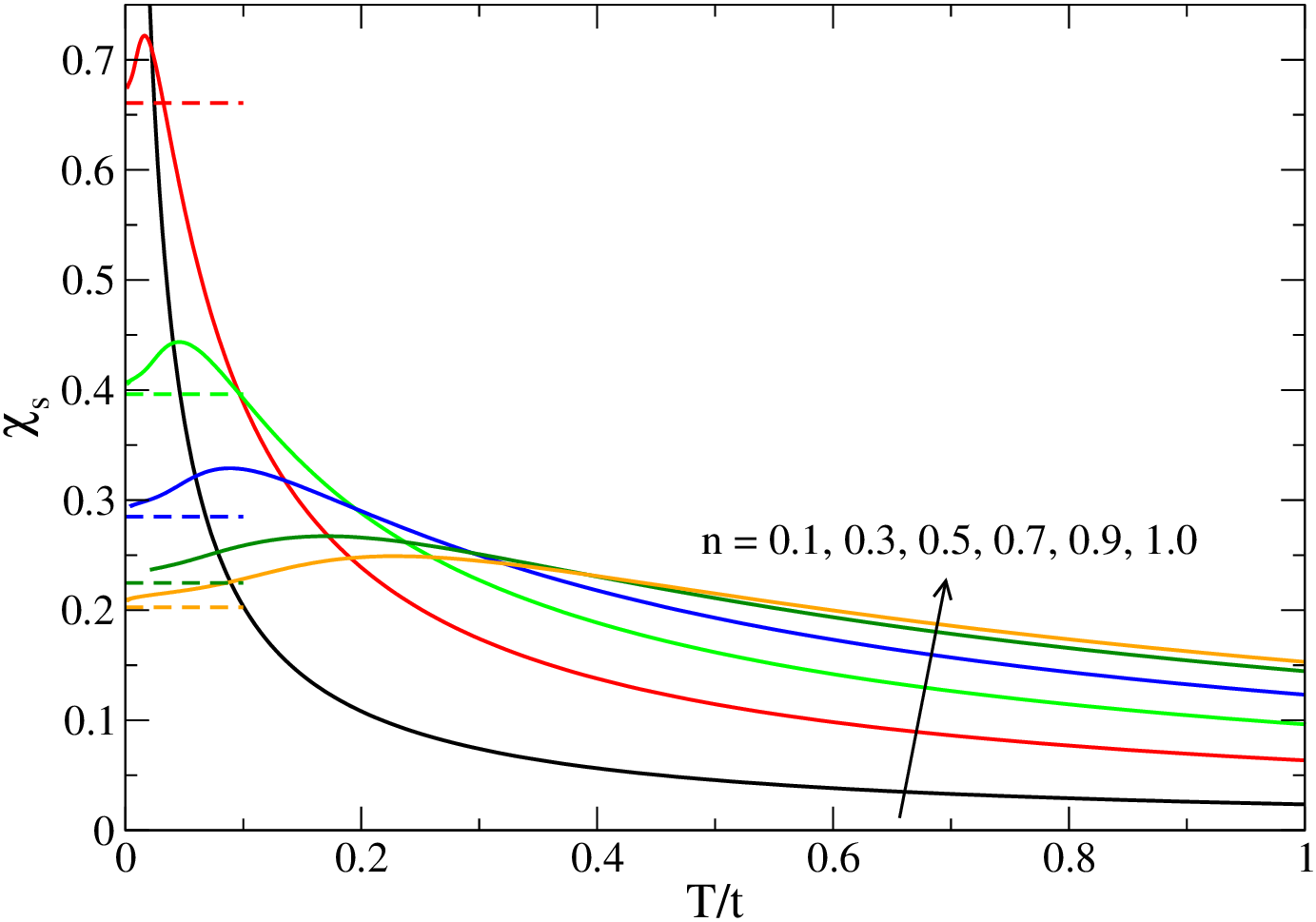}
\caption{(color online) Magnetic susceptibility $\chi_s$ for the integrable
  model. The dashed lines denote the zero temperature limit according to CFT.}
\label{BA.fig2}
\end{figure}

\begin{figure}
\includegraphics*[width=0.9\columnwidth]{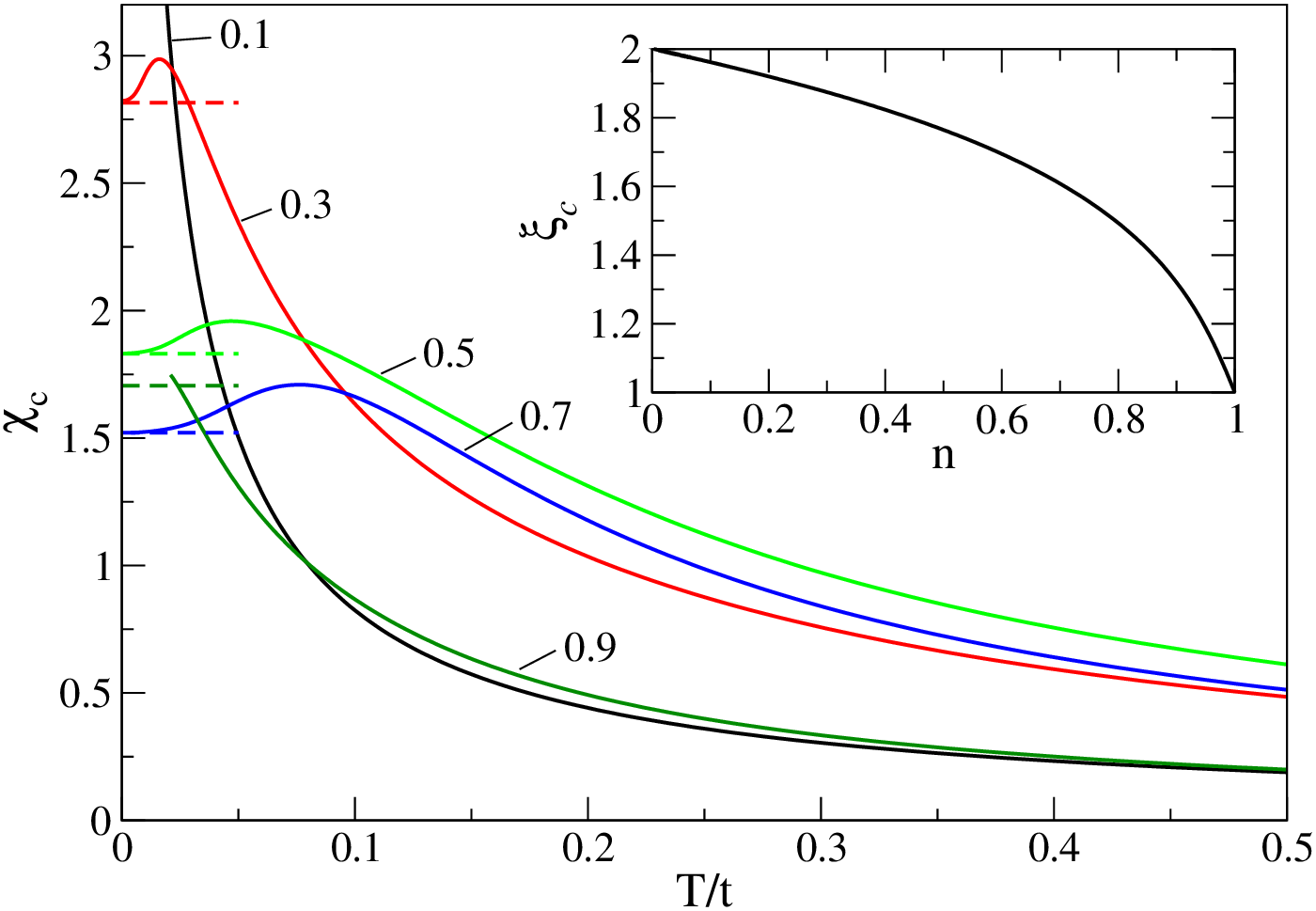}
\caption{(color online) Compressibility $\chi_c$ for the integrable model. The
  lines denote the zero temperature limit according to CFT. The inset shows
  the dressed charge as a function of density.}
\label{BA.fig3}
\end{figure}

The magnetic susceptibility $\chi_s$ as a function of temperature is shown in
Fig.~\ref{BA.fig2}.
From CFT we expect $\chi_s=1/\pi v_{so}$ at zero temperature. Because $v_{so}$
vanishes for $n\to 0$, the magnetic susceptibility diverges in this limit. For
$n= 1$, on the other hand, we have $v_{so}= J\pi/4$ so that $\chi_s=2/\pi^2$.
Note, however, that logarithmic corrections are expected at low temperatures
similar to the Heisenberg chain. Therefore the susceptibility will approach
the zero temperature limit predicted by CFT with infinite slope. This explains
why even at the lowest temperatures shown in Fig.~\ref{BA.fig2} the
susceptibility curves obtained by BA still deviate significantly from the zero
temperature limit. In Fig.~\ref{BA.fig3} we show the compressibility $\chi_c$
for various densities.
For the compressibility CFT predicts $\chi_c=K_c/(\pi v_c)$ where $K_c$ is the
Luttinger parameter of the charge sector. Using the BA we can calculate the so
called dressed charge\cite{Kawakami_SUN,FrahmSchadschneider,IzerginKorepin},
$\xi_c(Q)$, which is related to the Luttinger parameter by $K_c=\xi_c(Q)^2$.
The dressed charge as a function of density is shown in the inset of
Fig.~\ref{BA.fig3}. Because the charge velocity vanishes for $n\to 0$ and
$n\to 1$ (see Fig.~\ref{BA.fig1}) we have a diverging compressibility in both
limits. In addition, we notice that even at intermediate densities the
compressibility is large indicating that the integrable point is not that far
from a phase separated state. In general, phase separation is promoted by an
increasing ratio of $J/t$ as is well known in the $SU(2)$ $t-J$
model.\cite{OgataLuchini,HellbergMele} We will come back to this point in
Sec.~\ref{xy-model}.

\begin{figure}
\includegraphics*[width=0.9\columnwidth]{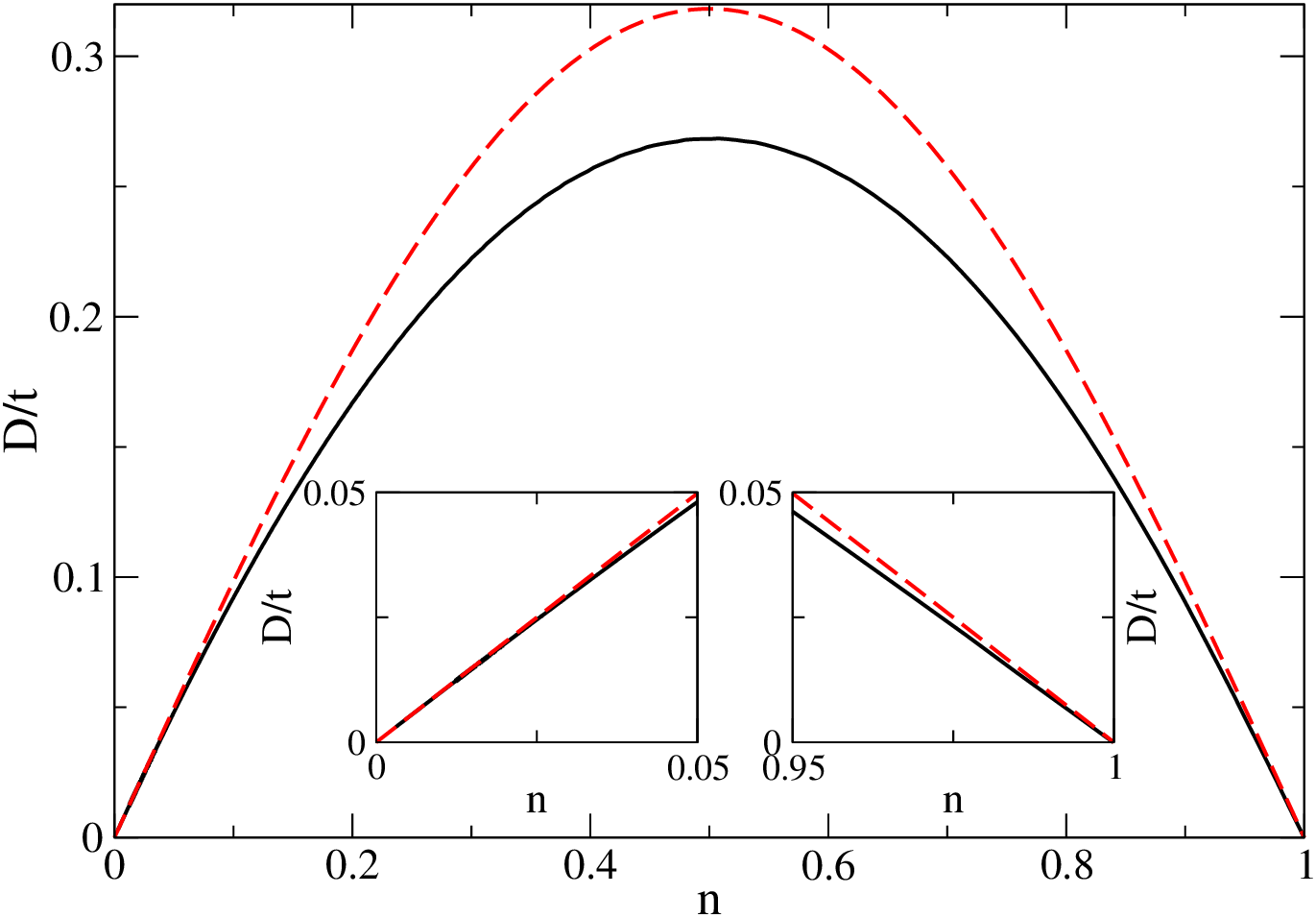}
\caption{(color online) Drude weight $D$ for the supersymmetric point $J=2t$
  (black solid line) and for $J=0$ (red dashed line), respectively. In the
  insets the regions of small and large densities are shown in detail.}
\label{BA.fig4}
\end{figure}

Finally, we want to consider the transport properties of the integrable model
at zero temperature. We can write the real part of the conductivity at zero
momentum as $\sigma'(\omega) = 2\pi D\delta(\omega)
+\sigma_{\mbox{reg}}(\omega)$ where $D$ is the {\it Drude weight} and
$\sigma_{\mbox{reg}}(\omega)$ the regular part. With the help of conformal
field theory and Bethe ansatz we find
\begin{equation}
\label{Drude}
D = \frac{K_cv_c}{2\pi} = \frac{\xi_c^2(Q)v_c}{2\pi} \; .
\end{equation}
Thus, the Drude weight follows directly from the dressed charge shown in the
inset Fig.~\ref{BA.fig3} and the charge velocity shown in the left inset of
Fig.~\ref{BA.fig1} and is depicted in Fig.~\ref{BA.fig4}.
$D$ vanishes for $n\to 0$ where the number of charge carriers vanishes, and
for $n\to 1$ where the Mott gap opens up. Quite surprisingly, the Drude weight
for the supersymmetric point appears to be symmetric around $n=1/2$ although
the Hamiltonian (\ref{intro2}) does not possess such a symmetry. To understand
this behavior it is instructive to study in addition the Drude weight for
$J=0$. In this case only the kinetic energy part of (\ref{intro2}) remains.
Then the spin and orbital indices do not matter and due to the projection
operators the model becomes equivalent to a {\it one-band, spinless} fermion
model. Note that the Mott gap at $n=1$ is therefore still incorporated. The
Drude weight can now be calculated from
\begin{equation}
\label{Drude2}
e(\Phi)-e(0) = D\Phi^2/N^2+\mathcal{O}(N^{-3}) 
\end{equation} 
with $e$ being the ground state energy per site, and $\Phi$ a field describing
a twist in the boundary conditions, $c_{j+N}=\e^{i\Phi}c_j$. We find
\begin{equation}
\label{Drude3}
D(J=0) =\frac{t}{\pi}\sin{\pi n}
\end{equation}  
which is shown as dashed curve in Fig.~\ref{BA.fig4}. Now the Drude weight is
indeed symmetric around $n=1/2$ because the spinless fermion model is
particle-hole symmetric. The comparison with the supersymmetric case near
$n\sim 0$ in the left inset of Fig.~\ref{BA.fig4} shows that the Drude weights
coincide in this limit. This is expected because the exchange interaction $J$
becomes irrelevant in the dilute limit. For $n\sim 1$ shown in the right inset
of Fig.~\ref{BA.fig4}, on the other hand, we see that the two curves do not
coincide. In this limit $J$ cannot be neglected. This means that $D$ for the
supersymmetric case is not symmetric around $n=1/2$ but the deviations from
this symmetry are very small. Similar observations have been made for the
one-band supersymmetric $t-J$ model by Kawakami and Yang,
Ref.~\onlinecite{KawakamiYang_Drude}.
\section{Density-matrix renormalization group}
\label{TMRG}
The density-matrix renormalization group applied to transfer matrices (TMRG)
is based on a mapping of a one-dimensional quantum system to a two-dimensional
classical one by means of a Trotter-Suzuki decomposition.  In the classical
model one direction is spatial whereas the other corresponds to the inverse
temperature. For the classical system a so called quantum transfer matrix
(QTM) is defined which evolves along the spatial direction. At any non-zero
temperature the QTM has the crucial property that its largest eigenvalue
$\Lambda_0$ is separated from the other eigenvalues by a finite gap. The
partition function of the system in the thermodynamic limit is therefore
determined by $\Lambda_0$ only, allowing it to perform this limit exactly. The
Trotter-Suzuki decomposition is discrete so that the transfer matrix has a
finite number of sites or local Boltzmann weights $M$. The temperature is
given by $T\sim (\epsilon M)^{-1}$ where $\epsilon$ is the discretization
parameter used in the Trotter-Suzuki decomposition. The algorithm starts at
some high-temperature value where $M$ is so small that the QTM can be
diagonalized exactly. Using a standard infinite-size DMRG algorithm, sites are
then added to the QTM leading to a successive lowering of the temperature. The
TMRG algorithm is described in detail in Refs.~\onlinecite{Peschel},
\onlinecite{GlockeKluemperSirker_Rev}, \onlinecite{SirkerKluemperEPL} and has
been applied to a number of one-dimensional systems such as frustrated and
dimerized spin chains,\cite{Raupach} the Kondo lattice
model,\cite{MutouShibata} the $t-J$
chain\cite{SirkerKluemperEPL,SirkerKluemperPRB} and ladder\cite{RiceTroyer} as
well as to the extended Hubbard model.\cite{GlockeSirker}

To obtain insight into the physical properties of the doped spin-orbital model
we will, in particular, be interested in the behavior of two-point correlation
functions. At finite temperatures we expect that any two-point correlation
function of a local operator $O(r)$ decays exponentially with distance $r$
\begin{equation}
\label{TMRG1}
\langle O(1) O(r)\rangle - \langle O(1)\rangle\langle O(r)\rangle = \sum_n M_n
\e^{-r/\xi_n}\e^{ik_n r} \; .
\end{equation}
Here $M_n$ is a matrix element, $\xi_n$ the correlation length, and $k_n$ the
corresponding wave vector. Note, that in the asymptotic expansion
(\ref{TMRG1}) infinitely many correlation lengths appear. Within the TMRG
algorithm, correlation lengths and corresponding wave vectors are determined
by next-leading eigenvalues $\Lambda_n$ of the QTM
\begin{equation}
\label{TMRG2}
\xi_n^{-1} = \ln\l|\frac{\Lambda_0}{\Lambda_n}\r| \quad , \quad k_n=\arg\l(\frac{\Lambda_n}{\Lambda_0}\r) \; .
\end{equation}
The long-distance behavior of the correlation function is then dominated by
the correlation length $\xi_\alpha$ belonging to the largest eigenvalue
$\Lambda_\alpha$ ($\alpha\neq 0$) with $M_\alpha\neq 0$. 

Apart from spin-spin (orbital-orbital) $\langle \vec{S}(1)\vec{S}(r)\rangle$ ($\langle
\vec{\tau}(1)\vec{\tau}(r)\rangle$) two-point correlation functions we are also
interested in pair correlation functions to investigate possible
superconducting instabilities. For the model (\ref{intro2}) we can define the
following singlet and triplet pair correlation functions
\begin{eqnarray}
\label{TMRG3}
G_{tt}(r) &=& \langle c_{\uparrow a}(r+1) c_{\uparrow a}(r)  c^\dagger_{\uparrow
  a}(2) c^\dagger_{\uparrow a}(1)\rangle \nn \\
G_{ss}(r) &=& \langle c_{\uparrow a}(r+1) c_{\downarrow b}(r)  c^\dagger_{\uparrow a}(2) c^\dagger_{\downarrow
  b}(1)\rangle  \\
G_{ts}(r) &=& \langle c_{\uparrow a}(r+1) c_{\uparrow b}(r)  c^\dagger_{\uparrow
  a}(2) c^\dagger_{\uparrow b}(1)\rangle \nn \\
G_{st}(r) &=& \langle c_{\uparrow a}(r+1) c_{\downarrow a}(r)  c^\dagger_{\uparrow
  a}(2) c^\dagger_{\downarrow a}(1)\rangle \nn \; .
\end{eqnarray}
Here $a,b$ denote the $\tau^z$-component and $\uparrow, \downarrow$ the
$S^z$-component. For each of these pair correlations an asymptotic expansion
(\ref{TMRG1}) exists and relation (\ref{TMRG2}) can be used to numerically
determine the corresponding leading correlation lengths.

\begin{figure}
\includegraphics*[width=0.9\columnwidth]{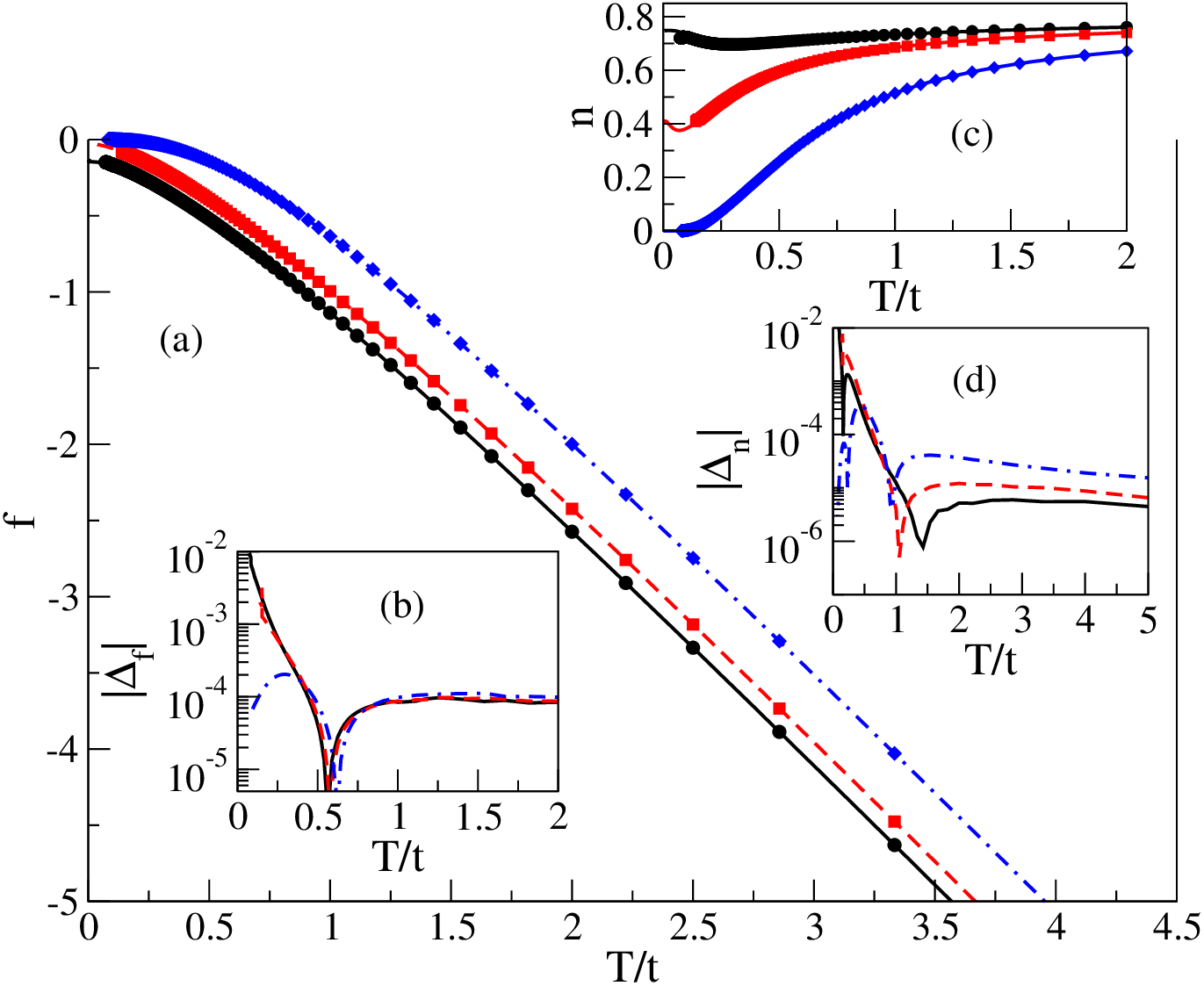}
\caption{(color online) (a): Free energy $f$ calculated by TMRG (symbols) with
  $N=360$ states kept compared to the exact solution (lines). The black
  circles (black solid lines) denote the result for $\mu=-1.7$, the red
  squares (red dashed lines) for $\mu=-1.9$, and the blue diamonds (blue
  dot-dashed lines) for $\mu=-2.5$, respectively. (b): Absolute error
  $|\Delta_f|$ of the TMRG results presented in (a). (c): TMRG results for the
  density compared to the exact solution with symbols and lines denoting the
  same chemical potentials as in (a). (d): Absolute error $|\Delta_n|$ of the
  TMRG results in (c).}
\label{TMRG.fig1}
\end{figure}

The discrete Trotter parameter $\epsilon$ leads to a systematic error in the
free energy of order $\epsilon^2$. In the calculations presented here we have
chosen $\epsilon =0.05$ so that this error is expected to be of the order
$10^{-3}-10^{-4}$. More important is the error due to the truncation of the
Hilbert space in each DMRG step. This error is difficult to estimate but
accumulates with each DMRG step finally leading to a breakdown of the numerics
at low temperatures. In the calculation presented here we will keep
$N=240-360$ states as basis for the truncated Hilbert space. To decide down to
which temperatures the TMRG is reliable we show results for the free energy
and density at the integrable point in Fig.~\ref{TMRG.fig1}.
It is important to note, that we perform the numerical calculations in a grand
canonical ensemble, i.e., we fix the chemical potential and not the particle
density. In particular for small doping levels it is, however, possible to
find a chemical potential so that the density depends only very weakly on
temperature as shown in Fig.~\ref{TMRG.fig1}(c). Note that for $T\to\infty$ we
always have $n\to 4/5$ because we have 5 states locally with one state
corresponding to the empty site. The absolute errors in the free
energies and in the densities stay smaller than $10^{-2}$ for temperatures down
to $T/t\sim 0.1$ as shown in Fig.~\ref{TMRG.fig1}(b) and (d),
respectively. This accuracy is completely sufficient to study the
thermodynamic properties of model (\ref{intro2}), and temperatures of the
order $T/t\sim 0.1$ are low enough to identify the ground state as well. 
\section{The $xy$-model}
\label{xy-model}
\begin{figure}
\includegraphics*[width=0.99\columnwidth]{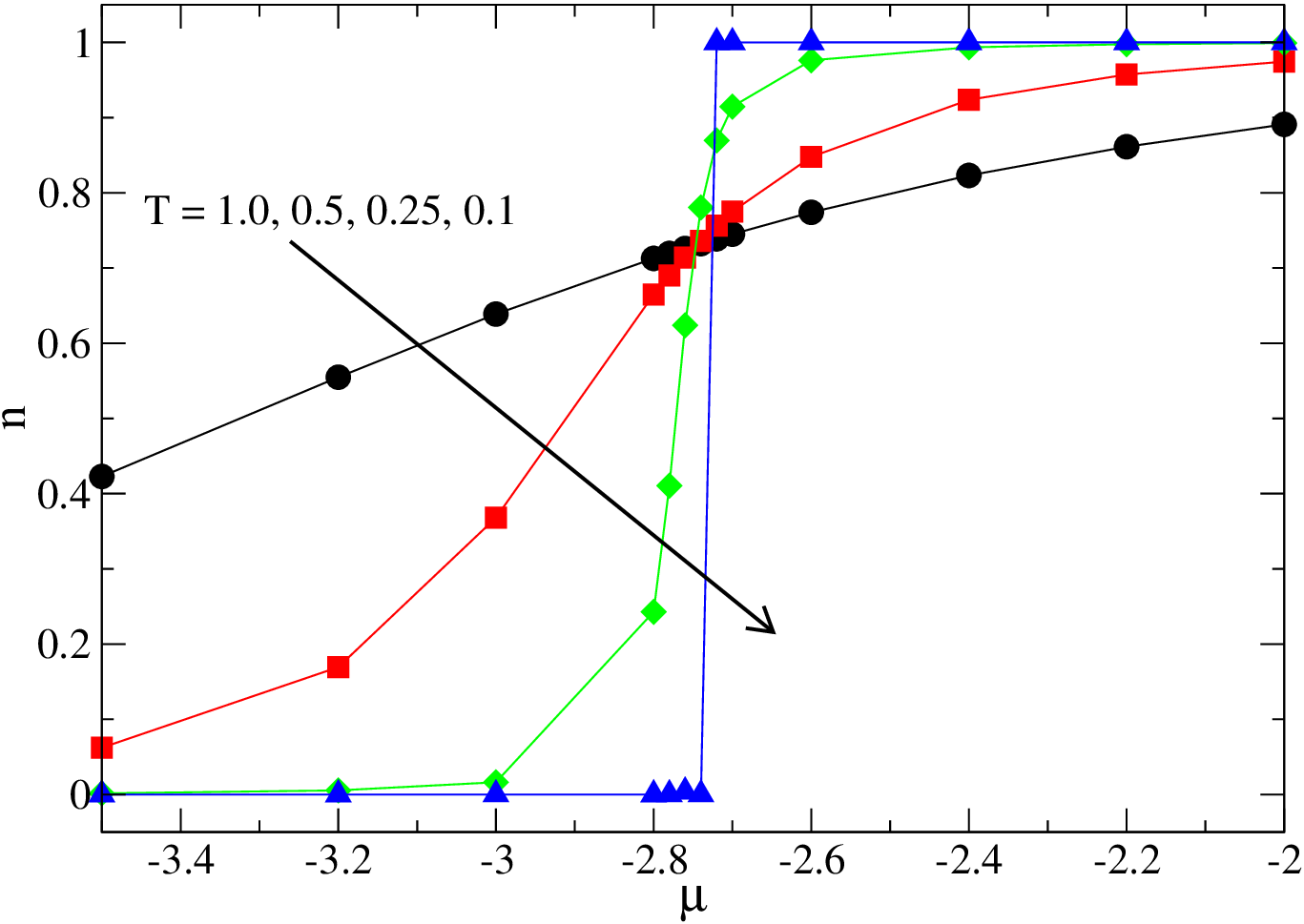}
\caption{(color online) The density $n$ as a function of chemical potential
  $\mu$ for different temperatures. Here $J/t=3$ and $x=y=1/4$. The lines are
  guides to the eye.}
\label{XY.fig1}
\end{figure}

In this section we want to investigate model (\ref{intro2}) away from the
integrable point. First, we want to demonstrate that the integrable point is
indeed already close to a state with phase separation. While keeping $x=y=1/4$
we now set $J/t=3$ and show in Fig.~\ref{XY.fig1} the density as a function of
chemical potential for various temperatures.  
For $T/t\to 0$ the density is zero for $\mu < -2.73$ and equal to one
otherwise. This means that the compressibility calculated at any fixed density
is divergent and the ground state therefore phase separated. The reason for
phase separation is obvious: For large $J/t$ the system tries to maximize its
magnetic exchange energy which is achieved by separating the particles from
the holes.

For all transition metal oxides we expect in general $J=4t^2/U<t$ which is
equivalent to $4t<U$. Exact values for $J/t$ depend on the considered compound
but values of $J/t\sim 0.3-0.5$ are typical. As a representative value we will
concentrate in the following on $J/t=0.5$. 
\subsection{The dimerized phase}
\label{dimerized}
\begin{figure}
\includegraphics*[width=0.99\columnwidth]{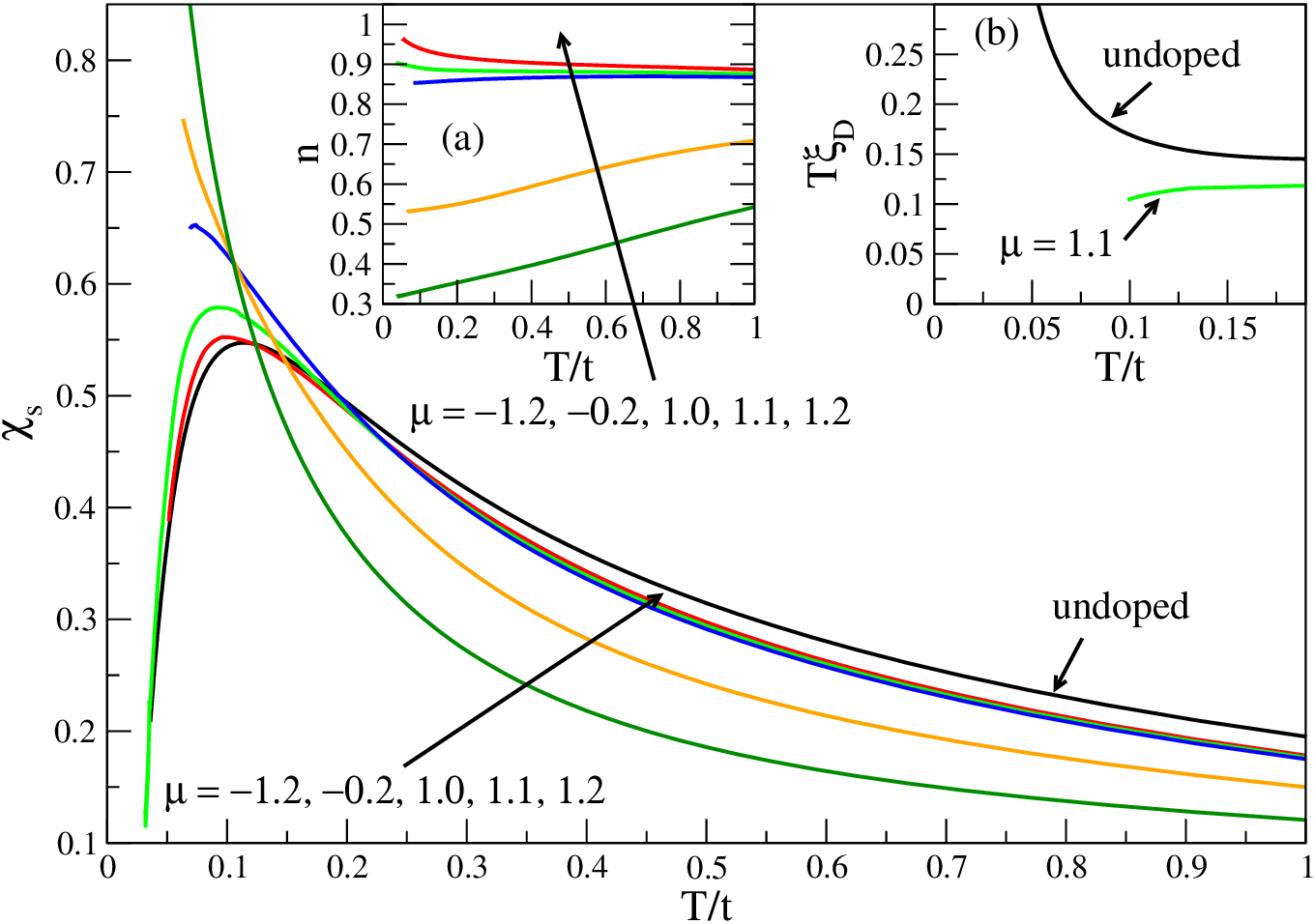}
\caption{(color online) Spin (orbital) susceptibility for $J/t=0.5$ and $x=y=1/2$ and
  different chemical potentials. Insets: (a) Corresponding densities as
  function of temperature. (b) Leading spin (orbital) dimer correlation length
  $\xi_D$. 
}
\label{XY.fig2}
\end{figure}

For $x=y=1/2$ the undoped model is in a dimerized phase and spin and orbital
excitations are gapped.\cite{ItoiQin,SirkerSu4} Spin-Peierls-type
instabilities are a generic feature of systems with coupled spin and orbital
degrees of freedom and have been investigated in more detail in
Refs.~\onlinecite{SirkerKhaliullin}, \onlinecite{SirkerHerzog}. Here we want
to study how dimer order and excitation gap evolve with doping. In
Fig.~\ref{XY.fig2} the magnetic susceptibility for various doping levels is
shown.
Note, that due to the $Z_2$ symmetry spin and orbital sectors are equivalent
and spin and orbital susceptibility therefore identical. In the undoped case a
spin gap $\Delta$ is clearly visible. In Ref.~\onlinecite{SirkerSu4} this gap
has been found to be of the order $\Delta = 0.090 \pm 0.005$. With increasing
hole concentration the spin gap becomes smaller but is still detectable
numerically at a chemical potential $\mu=1.0$ which corresponds to a density
$n=0.85$ at low temperatures. For larger doping levels the spin excitations
seem to become gapless so that the system apparently turns into a Luttinger
liquid. The long-range dimer order, on the other hand, seems to break down
immediately when holes are added to the system. In the case of algebraically
decaying correlations at zero temperature we expect the corresponding
correlation length to diverge as $\xi\sim 1/T$ whereas $\xi$ will diverge
stronger than $1/T$ in the case of true long-range order. In inset (b) of
Fig.~\ref{XY.fig2} we therefore show the leading spin (orbital) dimer
correlation length $\xi_D$ multiplied by $T$. In the undoped case the
divergence of $T\xi_D$ indicates that the ground state has indeed long-range
dimer order, whereas for a chemical potential $\mu = 1.1$, corresponding to
$n\sim 0.9$ at low temperatures, $T\xi_D$ is decreasing with temperature. Here
we expect $\xi_D$ to stay finite so that $T\xi_D\to 0$ for $T\to 0$ in accord
with the numerical data.
\subsection{Ferromagnetic/Antiferromagnetic-phase}
\label{FMAF}
\begin{figure}
\includegraphics*[width=0.99\columnwidth]{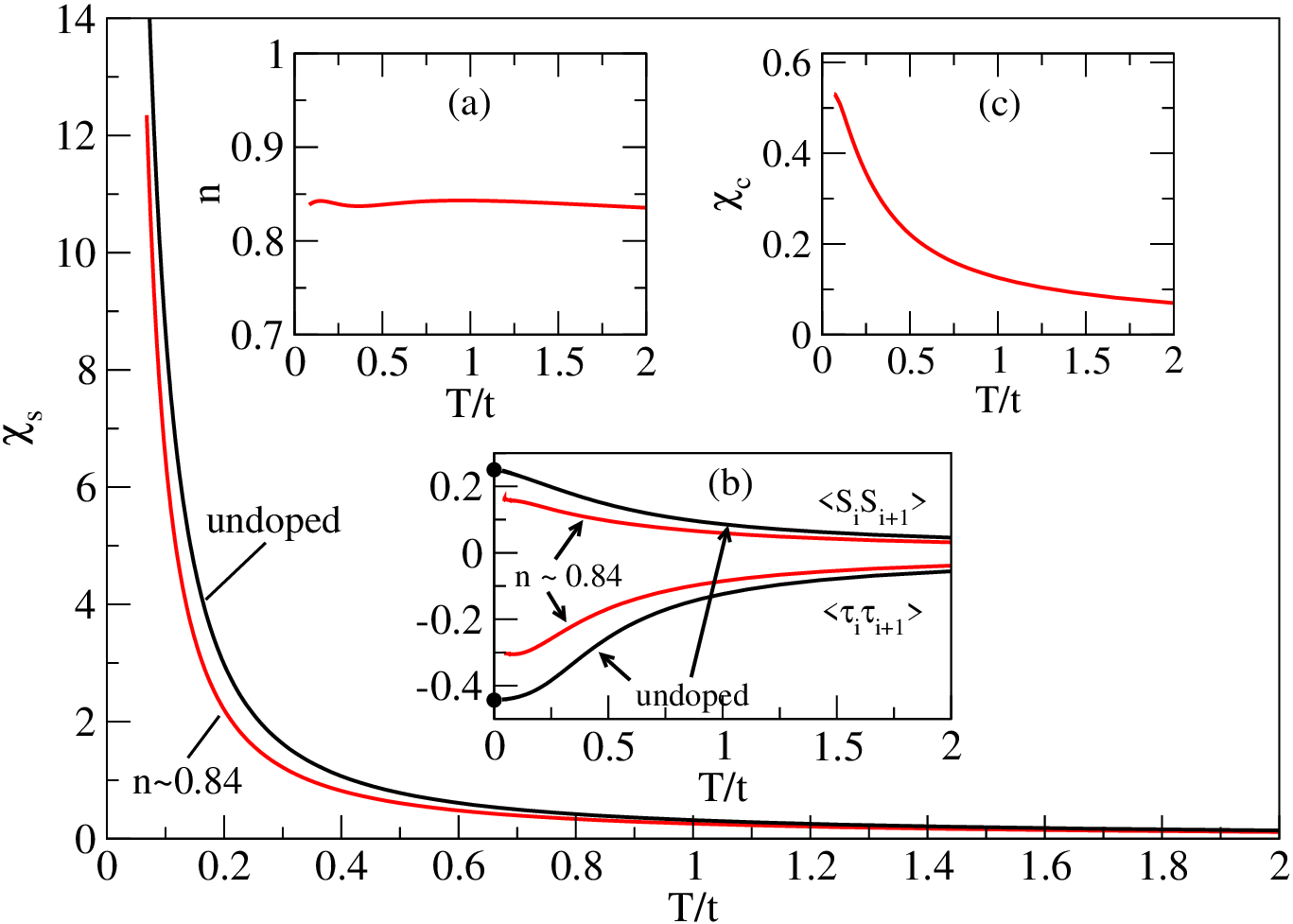}
\caption{(color online) TMRG results for $J/t=0.5$ with $x=1/2$ and $y=-1/2$.
  Main figure: Spin susceptibility in the undoped case and for $\mu=-0.1$ as
  function of temperature. Insets: (a) For $\mu=-0.1$ the density $n\sim 0.84$
  is almost independent of temperature. (b) Nearest-neighbor spin-spin and
  orbital-orbital expectation values for the undoped model and for $\mu=-0.1$
  ($n\sim 0.84$).  The circles on the $T=0$ axis denote $1/4$ and $-\ln 2
  +1/4$, respectively.  (c) Charge compressibility for $\mu=-0.1$.}
\label{XY.fig4}
\end{figure}

Here we want to consider $J/t=0.5$ with $x=0.5$, $y=-0.5$ where the
undoped model shows ferromagnetism in the spin sector and algebraically
decaying antiferromagnetic correlations in the orbital sector.\cite{ItoiQin}
As shown in inset (a) of Fig.~\ref{XY.fig4} we find that the density for a chemical
potential $\mu=-0.1$ is almost constant, $n\sim 0.84$, for temperatures $T\in
[0,2]$. This allows us to compare the undoped model directly with this slightly
doped case. 
In the main figure the spin susceptibility $\chi_s$ as a function of
temperature is shown. As expected, $\chi_s$ becomes suppressed with doping but
still diverges for $T\to 0$ indicating long-range ferromagnetic order in both
cases. In inset (b) of Fig.~\ref{XY.fig4} the nearest-neighbor spin-spin and
orbital-orbital expectation values are shown.  In the undoped case $\langle
\vec{S}_i \vec{S}_{i+1}\rangle \to 1/4$ for $T\to 0$ as expected for
ferromagnetic order. When the spins order ferromagnetically, then, according
to Hamiltonian (\ref{intro1}), we have an effective antiferromagnetic coupling
for the orbitals so that $\langle \vec{\tau}_i \vec{\tau}_{i+1}\rangle \to
-\ln 2+1/4$ for $T\to 0$. This is the value for an antiferromagnetic
Heisenberg chain known from Bethe ansatz. In the slightly doped case both the
ferromagnetic spin and the antiferromagnetic orbital correlations become
weaker. The charge compressibility shown in inset (c) of Fig.~\ref{XY.fig4}
is nonzero for $T\to 0$ indicating that the charge excitations are
gapless.

\begin{figure}
\includegraphics*[width=0.99\columnwidth]{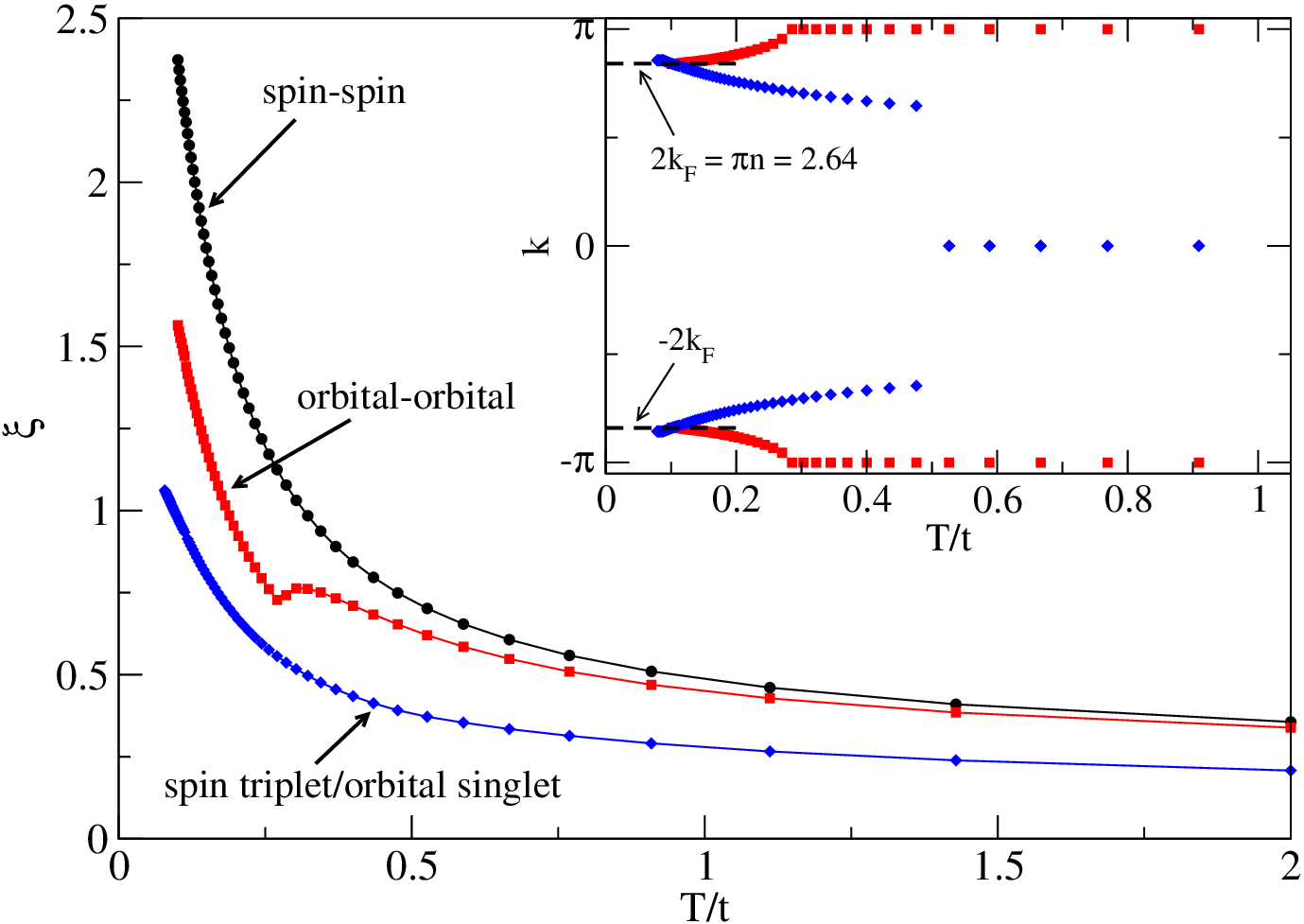}
\caption{(color online) Leading correlation lengths for $J/t=0.5$, $x=1/2$,
  $y=-1/2$. The leading spin-spin correlation length is non-oscillating
  ($k=0$) whereas the orbital-orbital and the pair correlations show
  incommensurate oscillations at low temperatures as depicted in the inset.}
\label{XY.fig5}
\end{figure}

This leads us to the question if the system has algebraically decaying pair
correlations and if so, which one of the pair correlations defined in
(\ref{TMRG3}) dominates. In Fig.~\ref{XY.fig5} we show some of the leading
correlation lengths.
Dominant is the spin-spin correlation length with wave vector $k=0$ which
diverges as $1/T$ indicative of the ferromagnetic order in the ground
state.\cite{TakahashiMSWT4} Next, we find a correlation length which belongs
to the asymptotic expansion of the orbital-orbital correlation function.
Interestingly, the associated wave vector is given by $k=\pi$ for $T/t\gtrsim
0.28$ but becomes incommensurate at lower temperatures. At $T/t\sim 0.28$ the
correlation length shows a cusp. The explanation for this cusp is as follows:
In the asymptotic expansion (\ref{TMRG1}) of the orbital-orbital correlation
function we have two correlation lengths with wave vector $k=\pi$. At $T/t\sim
0.28$ these two correlation lengths cross and the oscillations of the now
leading correlation length become incommensurate. For $T\to 0$ we find that
$k\to \pm 2k_F =\pm \pi n \approx \pm 2.64$ for $n\sim 0.84$. So at $T=0$ the
incommensurate oscillations just reflect the incommensurate filling of the
system. Similar crossover phenomena in the leading correlation length at
finite temperature have also been observed in the $xxz$ model in a magnetic
field \cite{KluemperScheeren} and in the one-band $t-J$ model at
incommensurate filling.\cite{SirkerKluemperEPL,SirkerKluemperPRB}

For the pair correlations defined in (\ref{TMRG3}) we find that the spin
triplet/orbital singlet correlation $G_{ts}$ has the largest correlation
length $\xi_{ts}$. At low temperatures we find $\xi_{ts}\sim 1/T$ indicating
that $G_{ts}$ will decay algebraically at zero temperature. The
associated wave vector is given by $k_{ts}=0$ for $T/t\gtrsim 0.5$ but becomes
incommensurate at lower temperatures. The numerical data seem to indicate that
for $T\to 0$ the oscillations become commensurate again with $k_{ts}=\pi$. In
the ground state this would mean that $G_{ts}(r)\sim (-1)^r/r^x$ with some
critical exponent $x$.

In the ferromagnetic/antiferromagnetic-phase considered in this section a
coupling between spin-orbital chains might therefore induce true long-range
triplet superconductivity.
Here triplet superconductivity arises from the coupling of the spins with the
orbital pseudospins. The degenerate orbitals order antiferromagnetically
leading to an effective ferromagnetic coupling for the spins.
\section{A one-dimensional version of a realistic model for cubic titanates}
\label{titanates}
In recent years a lot of interest has focused on the Mott insulator
LaTiO$_3$.\cite{KhaliullinMaekawa,Khaliullin_Rev} Here the octahedron of
oxygen ions surrounding each Ti$^{3+}$ is nearly perfect. This opens up the
possibility that the orbital degeneracy is not lifted by lattice distortions
and that the orbitals act as additional quantum degrees of freedom. Starting
from the ideal case of completely degenerate $t_{2g}$-orbitals one can derive
a superexchange model similar to (\ref{intro1}). Here the two orbitals
represented by the orbital pseudospin $\vec{\tau}$ depend on the bond
direction. Along the $c$-axis, for example, only the $t_{2g}$ levels of $xz$
and $yz$ symmetry are active and represented by $\vec{\tau}$, whereas
$\vec{\tau}$ stands for the $xy$ and $xz$ orbital if the bond is along the
$a$-axis. Having different orbital pairs active along each spatial direction
necessarily frustrates the one-dimensional physics discussed in the previous
sections of this paper and might lead to a liquid state with short range
$SU(4)$-type correlations.\cite{KhaliullinMaekawa} Nevertheless, a
directional, nematic state where the system makes full use of the orbital
quantum fluctuations say along the $c$-axis with active $xz$ and $yz$ orbitals
while the $xy$ orbital is empty, thus preventing orbital fluctuations in the
other two directions, might be close in energy and could possibly be realized
in LaTiO$_3$ under pressure.\cite{Khaliullin_Rev} An example, where such
orbital selection leading to a strongly directional spin-orbital state
probably happens is YVO$_3$. In this system, however, we have an effective
spin $S=1$.\cite{KeimerSirker,SirkerKhaliullin}

It is important to take the Hund's rule splitting of the virtual excited
states into account when deriving the superexchange Hamiltonian for LaTiO$_3$.
This leads, in particular, to an $xxz$-type anisotropy of the orbital sector,
i.e., contrary to (\ref{intro1}) the Hamiltonian no longer has a $SU(2)\times
SU(2)$ symmetry. If we consider the hole-doped case and add the hopping of the
holes to the superexchange Hamiltonian for LaTiO$_3$ given in
Ref.~\onlinecite{Khaliullin_Rev}, we obtain again a $t-J$ type Hamiltonian
which can be represented as
\begin{widetext}
\begin{eqnarray}
\label{real3}
H &=& t\sum_j\sum_{\sigma,\tau}\mathcal{P}\l\{
c^\dagger_{j,\sigma,\tau}c_{j+1,\sigma,\tau} + h.c. \r\}\mathcal{P} \\
 &+& 2J_\mathit{eff}\l[\l(\vec{S}_j\vec{S}_{j+1} +
x\r)\l(\vec{\tau}_j\vec{\tau}_{j+1} +\delta \tau^z_j\tau^z_{j+1}+
yn_jn_{j+1}\r)-\frac{z}{4}n_jn_{j+1}+\gamma \tau^z_j\tau^z_{j+1}\r] \nn
\end{eqnarray}
\end{widetext}
with parameters
\begin{eqnarray}
\label{real4}
&& J_\mathit{eff} = \frac{J}{2}(r_1+r_2) \quad , \quad
x=\frac{1}{4}+\frac{1}{2}\frac{r_1-r_2}{r_1+r_2} \nn \\
&& y =
\frac{1}{4}-\frac{1}{2}\frac{r_1-r_2}{r_1+r_2}+\frac{1}{6}\frac{r_3-r_2}{r_1+r_2}
\; , \; \delta = \frac{2}{3}\frac{r_3-r_2}{r_1+r_2} \nn \\
&& z= \frac{2}{3}\frac{r_1(5r_2+r_3)}{(r_1+r_2)^2} \quad , \quad \gamma =
\frac{2}{3}\frac{r_1(r_2-r_3)}{(r_1+r_2)^2} \; .
\end{eqnarray}
If we ignore the splitting of the virtually excited states due to Hund's
coupling we have $r_1=r_2=r_3=1$ so that the Hamiltonian (\ref{real4}) is
equivalent to the Hamiltonian (\ref{intro2}) with $x=y=1/4$ and
$J_\mathit{eff}=J$. For finite Hund's coupling $J_H$ we have
\begin{equation}
\label{real5}
r_1=\frac{1}{1-3\eta}\; , \; r_2=\frac{1}{1-\eta}\; , \; r_3=\frac{1}{1+2\eta}
\end{equation}
where $\eta=J_H/U$. 

\begin{figure}
\includegraphics*[width=0.99\columnwidth]{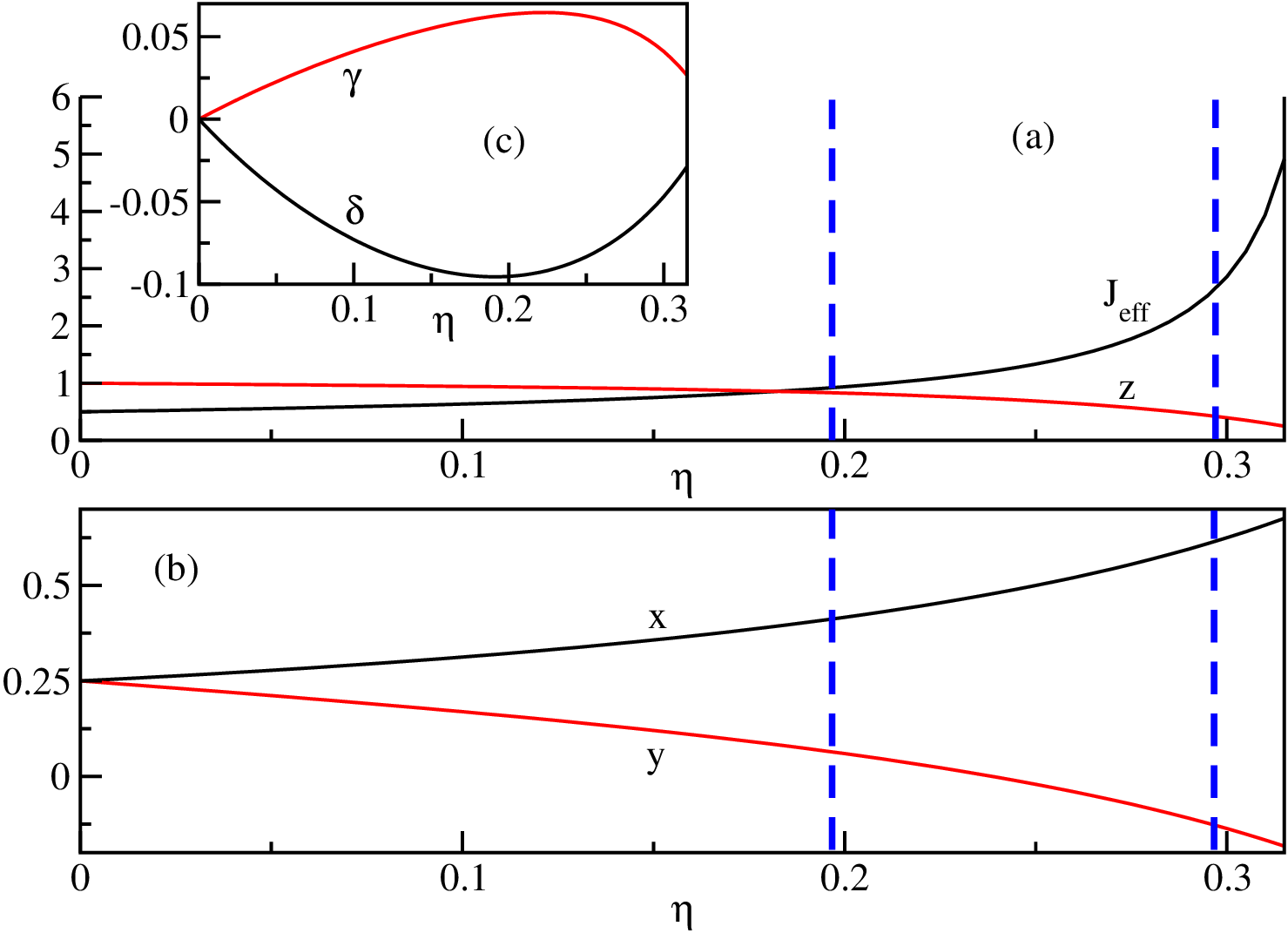}
\caption{(color online) (a) Parameters $J_\mathit{eff}$ for $J=0.5$ and $z$ as
  defined in Eq.~(\ref{real4}) as a function of $\eta=J_H/U$. (b) Parameters
  $x$ and $y$ as a function of $\eta$. (c) Parameters $\delta$ and $\gamma$
  related to the $xxz$-anisotropy of the orbital sector as a function of
  $\eta$. The dashed blue lines in (a) and (b) indicate the approximate values
  for the phase transitions described in the text.}
\label{real.fig1}
\end{figure}

From optical data and first principle calculations for LaTiO$_3$ one finds the
approximate values for the hopping amplitude $t\sim 0.3$ eV, the onsite
Coulomb repulsion $U\sim 2.8$ eV, and the Hund's rule coupling $J_H\sim 0.6$
eV.\cite{Khaliullin_Rev} For the model (\ref{real3}) this means that
$J=4t^2/U\sim 0.13$ eV, $J/t=4t/U\sim 0.43$, and $\eta=J_H/U\sim 0.21$. It is
therefore reasonable to set again $J/t=0.5$ as in the previous section. The
parameters (\ref{real4}) as a function of $\eta$ for this value of $J/t$ are
shown in Fig.~\ref{real.fig1}. 
Although $\delta$ and $\gamma$ become nonzero for finite $\eta$ thus
destroying the $SU(2)$ symmetry of the orbital sector, their values remain
small in the physical regime for $\eta$ depicted in Fig.~\ref{real.fig1}. It
is therefore indeed reasonable to neglect this anisotropy in a first
approximation as has been done in the previous section. Furthermore, we also
find that $z\sim 1$ at least up to $\eta\sim 0.2$ so that the variation in $z$
can also be neglected. We are then back to Hamiltonian (\ref{intro2}) with $J$
replaced by an effective superexchange scale $J_\mathit{eff}$ and with $x$ and $y$
being functions of the single parameter $\eta$ only. From this observation we
can infer the basic properties of this model: For $\eta\lesssim 0.2$ the model
will be in a ``rescaled $SU(4)$ phase'', i.e., a phase where the same field
theory as at the $SU(4)$ symmetric point describes the low-energy properties
but with spin and orbital velocities which are rescaled and no longer
equivalent. Strictly speaking this is only correct without orbital anisotropy
($\delta=\gamma =0$). Depending on the spin order, the orbital anisotropy
might become Ising-like so that the orbital excitations become gapped.
However, even if this happens the orbital gap caused by this mechanism will be
extremely small. For $\eta\gtrsim 0.2$ we expect to enter a phase with
ferromagnetically ordered spins and antiferromagnetic correlations in the
orbital sector. At the same time $J_\mathit{eff}$ increases with increasing $\eta$ so
that we expect a phase separated state if $\eta\gtrsim 0.3$ in the doped case.

\begin{figure}
\includegraphics*[width=0.99\columnwidth]{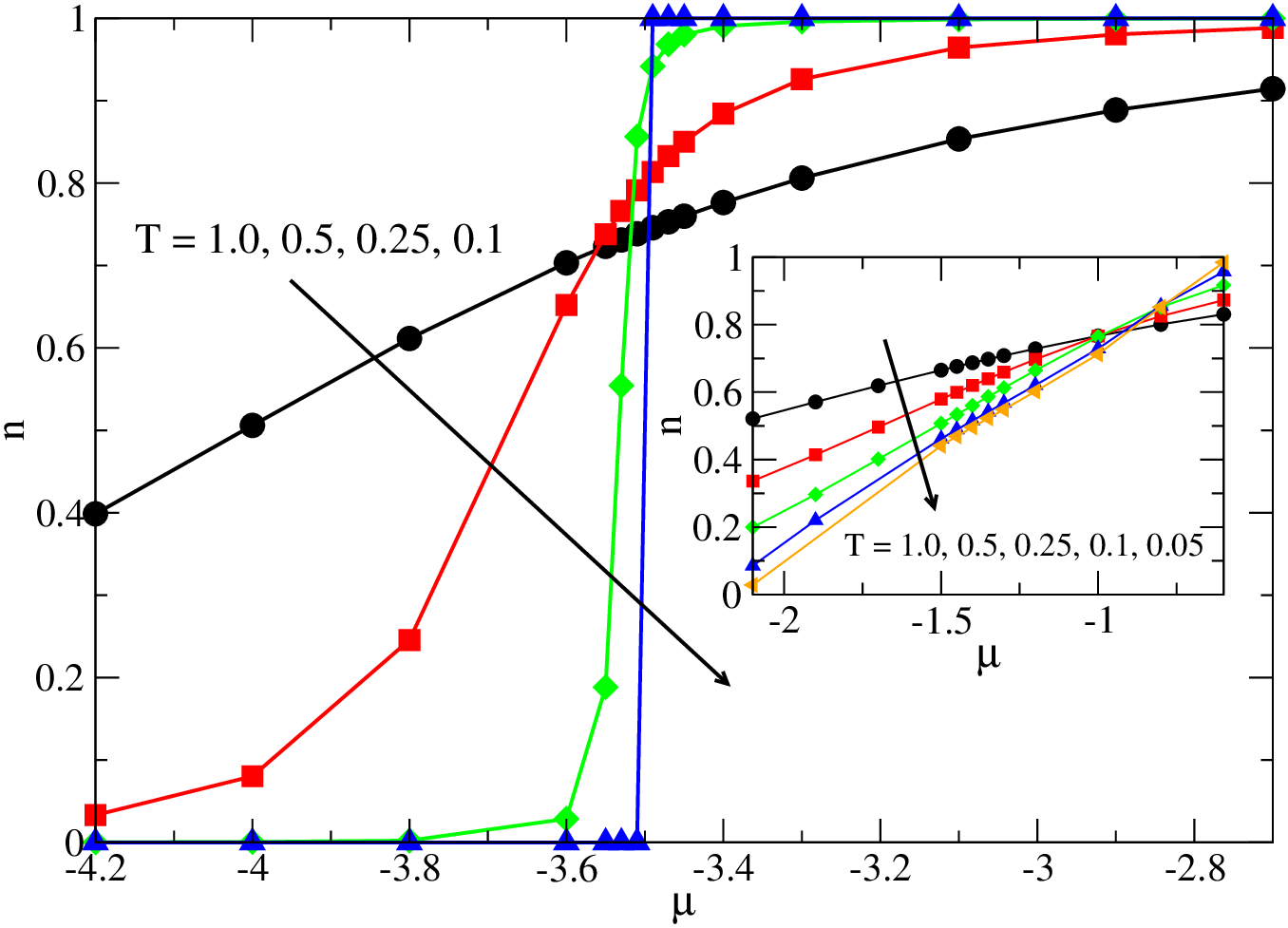}
\caption{(color online) Density as a function of the chemical potential for
  $J/t=0.5$ and $\eta=0.3$. The inset shows the same for $\eta=0.25$. The
  lines are guides to the eye.}
\label{real.fig2}
\end{figure}

In the following, we present numerical results for the full model
(\ref{real3}) with $J/t=0.5$ and the parameters as given in
Eqs.~(\ref{real4},\ref{real5}). In Fig.~\ref{real.fig2} the density as a
function of chemical potential is shown for $\eta=0.3$ and $\eta=0.25$.
As in Fig.~\ref{XY.fig1} we see that the density for $\eta=0.3$ and $T/t\to 0$
jumps from zero to one. Here the jump occurs at a chemical potential
$\mu\approx -3.5$. Again this indicates a diverging compressibility in a
canonical ensemble for all densities and confirms the expected phase
separation at large $\eta$. For $\eta=0.25$ (shown the in the inset of
Fig.~\ref{real.fig2}), on the other hand, the ground state is not phase separated.

\begin{figure}
\includegraphics*[width=0.99\columnwidth]{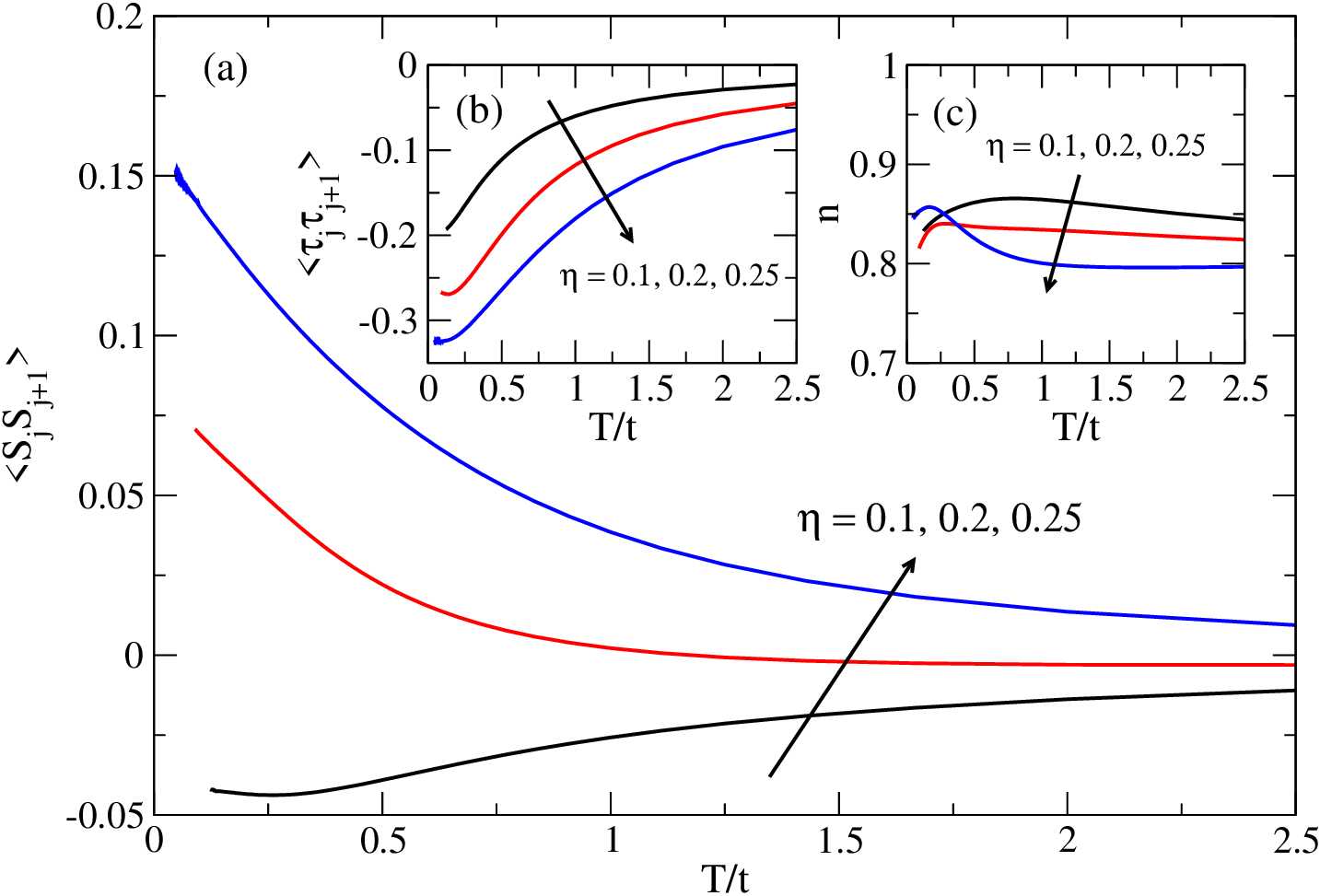}
\caption{(color online) TMRG results for model (\ref{real3}) with $J=0.5$ and
  $\eta=0.1,\, 0.2, 0.25$. The chemical potential is set to $\mu = 0.6$ for
  $\eta=0.1$, $\mu=0.0$ for $\eta=0.2$, and $\mu=-0.8$ for $\eta=0.25$,
  respectively.  (a) Nearest-neighbor spin-spin correlation
  $\langle\vec{S}_j\vec{S}_{j+1}\rangle$, (b) nearest-neighbor
  orbital-orbital correlation $\langle\vec{\tau}_j\vec{\tau}_{j+1}\rangle$,
  and (c) the density as a function of temperature. Note, that for the
  chemical potentials chosen here the densities depend only weakly on
  temperature, $n\sim 0.8-0.85$.}
\label{real.fig3}
\end{figure}

\begin{figure}
\includegraphics*[width=0.99\columnwidth]{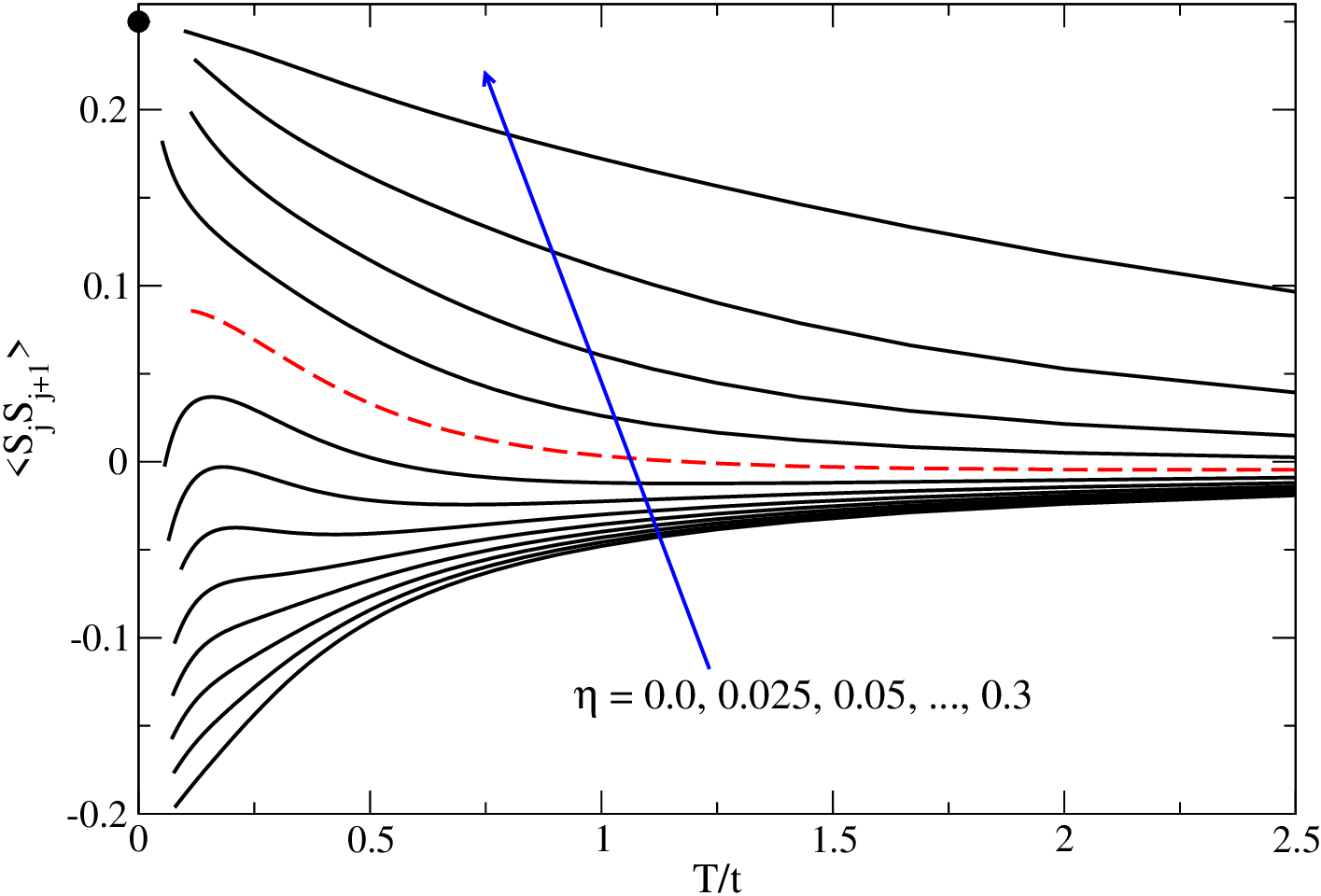}
\caption{(color online) TMRG results for the nearest-neighbor correlation
  function $\langle\vec{S}_j\vec{S}_{j+1}\rangle$ in the undoped case for
  $J=0.5$ and different $\eta$. The red dashed curve denotes the result for
  $\eta=0.2$.  The spins are fully polarized in the ground state,
  $\langle\vec{S}_j\vec{S}_{j+1}\rangle = 1/4$ (black dot), if $\eta\gtrsim
  0.2$.}
\label{real.fig4}
\end{figure}

Next, we consider the slightly doped case, $n\sim 0.8-0.85$, for different
parameters $\eta$. The nearest-neighbor correlation functions
$\langle\vec{S}_j\vec{S}_{j+1}\rangle$ and
$\langle\vec{\tau}_j\vec{\tau}_{j+1}\rangle$ presented in Fig.~\ref{real.fig3}
show clearly that the spin correlations are antiferromagnetic for $\eta=0.1$
and ferromagnetic for $\eta=0.2$ and $\eta=0.25$. 
The orbital correlations, on the other hand, are antiferromagnetic in all
three cases. To fix the critical value for $\eta$ where the phase transition
occurs, we consider in Fig.~\ref{real.fig4} the nearest-neighbor spin
correlation for the undoped model ($n=1$) as a function of temperature $T$ and
Hund's coupling $\eta$.
The data show that the spins in the ground state are fully polarized if $\eta
> \eta_c\approx 0.2$. For $\eta<\eta_c$ the spin correlations are
antiferromagnetic in the ground state. In this case, however,
$\langle\vec{S}_j\vec{S}_{j+1}\rangle$ can be non-monotonic as a function of
temperature and even larger than zero in a certain temperature range if $\eta$
is close to $\eta_c$. The expectation value for
$\langle\vec{S}_j\vec{S}_{j+1}\rangle$ jumps at zero temperature from some
$\eta$-dependent value for $\eta < \eta_c$ to $1/4$ for $\eta > \eta_c$. The
phase transition driven by $\eta$ is therefore first order. Very
interestingly, the phase transition occurs at a value for $\eta$ which is
close to the one expected for LaTiO$_3$. In addition to the frustration of the
orbital sector in this compound due to different orbital pairs being active
along each direction, this closeness to the phase transition might also be
important to understand the peculiar physics of LaTiO$_3$. It might, in
particular, be a contributing factor to the smallness of the ordered moment in
the G-type antiferromagnetic structure.\cite{KeimerCasa}
Note, however, that also other factors like the Ti-O-Ti bond angle are very
important in determining whether the spins order ferro- or
antiferromagnetically. The different magnetic properties of YTiO$_3$
(ferromagnetic spin order) and LaTiO$_3$ (G-type antiferromagnetic order), for
example, have been ascribed to a small variation in this
angle.\cite{khaliullinOkamoto2} Nevertheless, this variation in bond angle can
be mimicked to a certain degree by increasing $\eta$ so that the phase
transition situated at $\eta_c\approx 0.2$ in the one-dimensional model is
indeed important to understand the physics of the cubic titanates.

\begin{figure}
\includegraphics*[width=0.99\columnwidth]{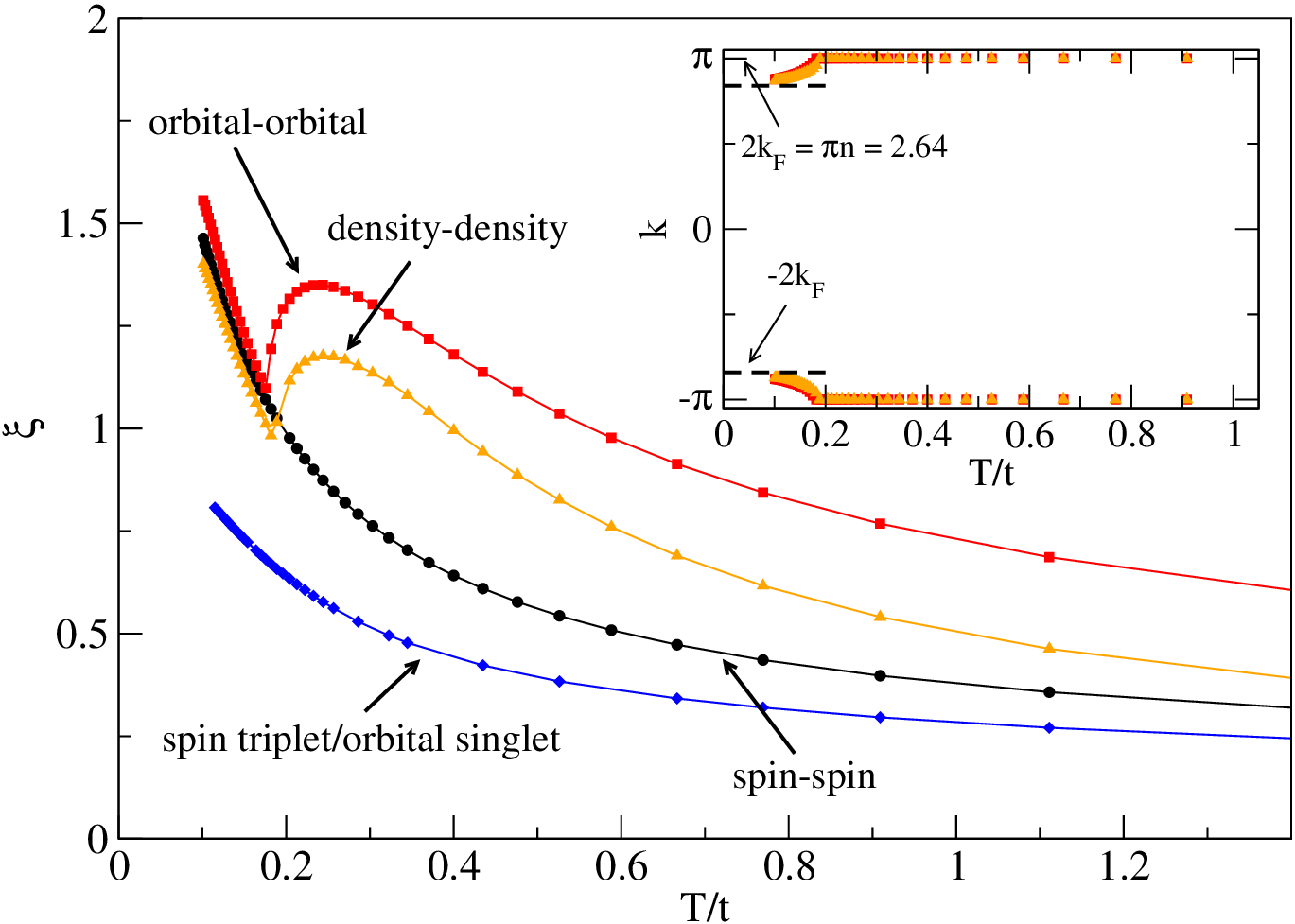}
\caption{(color online) Leading correlation lengths for $J=0.5$, $\eta=0.25$
  and a chemical potential $\mu=-0.8$ corresponding to a density $n\sim 0.84$
  at low temperatures (see Fig.~\ref{real.fig3}(c)). The orbital-orbital and
  the density-density correlations show incommensurate oscillations at low
  temperatures as depicted in the inset whereas the spin-spin and the spin
  triplet/orbital singlet correlations are both non-oscillating.}
\label{real.fig5}
\end{figure}

Finally, we again want to study possible pairing instabilities in the
slightly doped case. Here we concentrate on $\eta=0.25$ with $J=0.5$
and $\mu=-0.8$ as in Fig.~\ref{real.fig3}. Results for the leading
correlation lengths in this case are presented in
Fig.~\ref{real.fig5}.
We find that the spin-spin, orbital-orbital, density-density, as well as the
spin triplet/orbital singlet correlation lengths all diverge as $1/T$ for
$T\to 0$ indicating that these correlations will decay algebraically at zero
temperature. As in Sec.~\ref{FMAF} we find that the oscillations of the
orbital-orbital and the density-density correlation become incommensurate in
the low temperature regime. Again $k\to \pm 2k_F =\pm \pi n \approx \pm 2.64$
for $n\sim 0.84$ reflecting the incommensurate filling of the system. The spin
triplet/orbital singlet correlation length is larger than any of the other
correlation lengths associated with the pair correlations defined in
(\ref{TMRG3}). If superconductivity can be stabilized at all in a possible
nematic phase of LaTiO$_3$ or YTiO$_3$ there is therefore the possibility that
it will be of triplet character.
\section{Summary and Conclusions}
\label{Conc}
In this paper we studied the thermodynamic properties of hole-doped
one-dimensional Mott insulators with orbital degrees of freedom. We described
such systems in terms of generalized (multi-band) $t-J$ models. Neglecting the
Hund's rule splitting of the virtually excited states we were led to a model
which is integrable for one specific value of the ratio $J/t$. At this point
the model becomes $SU(4|1)$ symmetric (graded $SU(5)$ symmetry) and belongs to
the so called Uimin-Sutherland class of models.\cite{Sutherland} The
integrability at this particular point is analogous to the integrability of
the usual $t-J$ model at the supersymmetric point. Ground-state properties of
the $SU(4|1)$ symmetric spin-orbital model have been first investigated using
Bethe ansatz by Schlottmann and
Kawakami.\cite{Schlottmann_SU5_1,Schlottmann_SU5_2,Kawakami_SUN} Here we
presented a set of nonlinear integral equations also based on the Bethe ansatz
which allowed us to study the thermodynamics. Using conformal field theory we
have been able to connect our new results for the thermodynamics at low
temperatures with the results for the ground state and the elementary
excitations obtained by Schlottmann and Kawakami.

In the second part of the paper we used the density matrix renormalization
group applied to transfer matrices (TMRG) to study the thermodynamics of the
two-band $t-J$ model away from the integrable point. By comparing with Bethe
ansatz results at the integrable point we first demonstrated that the obtained
numerical results are accurate down to low temperatures $T/t\sim 0.05$. For
large values of $J/t$ we then showed that the ground state becomes phase
separated. Next, we studied the effects of hole doping on various phases of
the undoped model. If the ground state of the undoped model is dimerized then
the associated spin gap persists up to relatively large hole concentrations.
The long-range nature of the dimer order, however, seems to break down
immediately upon hole doping. Starting from the phase of the undoped model
where the spins are fully ferromagnetically polarized and the orbitals show
antiferromagnetic correlations we found that upon hole doping the spin
triplet/orbital singlet pair correlation dominates among the various possible
pair correlation functions. This correlation function will decay algebraically
at zero temperature so that interchain couplings might stabilize true triplet
superconductivity in this phase.

In the last part we used the TMRG algorithm to study the effects of hole
doping for a one-dimensional version of a realistic model for cubic titanates.
The model can be written in a form making it very similar to the one
considered earlier. Previously independent parameters, however, now become
functions of the ratio $\eta$ of Hund's coupling $J_H$ and the onsite Coulomb
repulsion $U$, and there is an additional $xxz$-type anisotropy of the orbital
sector. The effective superexchange coupling is a function of $\eta$ as well
and increases monotonically with increasing $\eta$. For $\eta\gtrsim 0.3$ this
leads to a phase separated ground state. In addition, we find a phase
transition at $\eta_c\approx 0.2$ between a state with antiferromagnetically
correlated spins ($\eta<\eta_c$) and a state with fully ferromagnetically
polarized spins ($\eta>\eta_c$). The phase transition is first order. For
$\eta>\eta_c$ the dominating pair correlation is again of spin triplet/orbital
singlet character and might lead to a true superconducting instability if
interchain couplings are present. Interestingly, realistic $\eta$-values for
LaTiO$_3$ are close to $\eta_c$. Although a purely one-dimensional model is
not appropriate for this compound, the nature of local correlations might be
correctly captured and $\eta$ being close to $\eta_c$ might contribute to the
peculiar properties of this compound, in particular, to the extremely small
G-type magnetic moment.

\begin{acknowledgments}
JS thanks G.~Khaliullin and P.~Horsch for valuable discussions. JD
gratefully acknowledges financial support by the Volkswagen Foundation
and by the DFG through Graduiertenkolleg 1052.
\end{acknowledgments}

\appendix
\section{Nonlinear integral equations}
\label{appA}
The integrable $SU(4|1)$ model admits the calculation of exact results for the
thermodynamics. Here the largest eigenvalue of the QTM can be obtained by
Bethe ansatz.\cite{KluemperWehnerZittartz} The number of Bethe ansatz
equations, however, diverges in the limit $M \to \infty$. It is thus necessary
to encode the Bethe ansatz equations into an alternative form for which the
limit can be taken analytically. This can be done by defining suitable
auxiliary functions in the spirit of Refs.~\onlinecite{Kluemper1,Kluemper2},
which are shown to be determined by a closed set of only finitely many coupled
nonlinear integral equations~(NLIEs).

The rigorous derivation depends on the explicit knowledge of the
auxiliary functions in terms of the Bethe ansatz roots for finite
$M$. Unfortunately these are unknown for the $SU(4|1)$ model. Yet we
are able to conjecture the complete set of coupled NLIEs by
generalizing the structure that has been found for two closely related
models in the Uimin-Sutherland class, namely the $SU(2|1)$ and the
$SU(4)$ models.\cite{JuettnerKluemper,DamerauKluemper} For the
$SU(4|1)$ model we thus expect a total number of 15 coupled NLIEs,
exactly one more than for the $SU(4)$ model. Their structure should be
given by
\begin{multline}
  \ln\rep{b}{a}_j(x) = -\frac{t \rep{V}{a}(x) + \rep{c}{a}_j}{T}\\
  -\sum_{b=1}^{4} \sum_{k=1}^{\binom{4}{b}} \int_{-\infty}^\infty
  \rep{K}{a,b}_{j, k}(x - y) \ln\rep{B}{b}_k(y)\,\frac{\mathrm d y}{2
    \pi}
\end{multline}
where $\rep{B}{a}_j(x) = \rep{b}{a}_j(x) + 1$ are the unknown
auxiliary functions. The free energy is obtained from these functions
via
\begin{equation}
  f = -T \sum_{a=1}^4 \sum_{j=1}^{\binom{4}{a}} \int_{-\infty}^\infty
  \rep{V}{a}(y) \ln\rep{B}{a}_j(y) \, \frac{\mathrm d y}{2 \pi} \; .
\end{equation}
Since the NLIEs must both reproduce the known results for the $SU(4)$
model in the limit $n \to 1$ ($\mu \to \infty$) and yield the correct
zero-temperature limit, it is possible to fix the driving terms and
kernel functions. We find
\begin{equation}
  \rep{V}{a}(x) = \frac{4 a}{4 x^2 + a^2}
\end{equation}
and the constants
\begin{align}
  \rep{c}{1}_1 &= \rep{c}{1}_2 = -2 t - \mu - h / 2 \notag\\
  \rep{c}{1}_3 &= \rep{c}{1}_4 = -2 t - \mu + h / 2 \notag\displaybreak[0]\\
  \rep{c}{2}_1 &= -4 t - 2 \mu - h \notag\\
  \rep{c}{2}_2 &= \rep{c}{2}_3 = \rep{c}{2}_4 = \rep{c}{2}_5 = -4 t -
  2 \mu \notag\\
  \rep{c}{2}_6 &= -4 t - 2 \mu + h \notag\displaybreak[0]\\
  \rep{c}{3}_1 &= \rep{c}{3}_2 = -6 t - 3 \mu - h / 2 \notag\\
  \rep{c}{3}_3 &= \rep{c}{3}_4 = -6 t - 3 \mu + h / 2 \notag\\
  \rep{c}{4}_1 &= -8 t - 4 \mu \; .
\end{align}
The kernel functions $\rep{K}{a,b}_{j,k}(x)$ for $a,b =
1,2,3$ are similar to those of the $SU(4)$ model (see
Ref.~\onlinecite{DamerauKluemper} eqs.~(33)--(35)), but where the
common functions $\rep{\widehat K}{a,b}_{[4]}(k)$ are replaced by
\begin{equation}
  \rep{\widehat{\mathcal K}}{a,b}(k) = \e^{(1 - b) |k| /
    2} \, \frac{\sinh(a k / 2)}{\sinh(k / 2)} -
  \delta_{a,b} \; .
\end{equation}
The remaining kernel functions are
\begin{equation}
  \rep{K}{4,a}_{1,j}(x) = \rep{K}{a,4}_{j,1}(x) = \int_{-\infty}^\infty
  \rep{\widehat{\mathcal K}}{a,4}(k) \e^{\im k x}\,\mathrm d k \; .
\end{equation}

The set of NLIEs can easily be solved numerically by iteration yielding high
accuracy over the whole parameter range. The validity of the results has been
checked by comparing our specific heat data to the high-temperature expansion
rigorously derived in Ref.~\onlinecite{Tsuboi} on the basis of an alternative set
of NLIEs. Moreover, the results agree in the low-temperature limit with CFT
and over the whole temperature range with numerical TMRG calculations as shown
in this article.


\end{document}